\documentclass[aps,prd,superscriptaddress,nofootinbib,amsmath,amsfonts,preprintnumbers,groupedaddress,showpacs,9pt,english]{revtex4-1}
\usepackage{amsmath}
\usepackage{amssymb}
\usepackage{babel}
\usepackage{wrapfig}
\usepackage{cancel}
\usepackage{array}   

\usepackage{graphicx}
\graphicspath{{figures/}}
\usepackage{booktabs}
\usepackage{amsmath,amssymb}  
\usepackage{mathrsfs}         
\usepackage[colorlinks,citecolor=blue,urlcolor=blue,linkcolor=blue]{hyperref}
\usepackage[figtopcap]{subfigure}
\usepackage{color}
\usepackage{relsize,exscale}
\makeatletter
\newcommand{\beq}{\begin{equation}}
\newcommand{\eeq}{\end{equation}}
\newcommand{\bea}{\begin{eqnarray}}
\newcommand{\eea}{\end{eqnarray}}

\newcommand{\comm}[1]{}

\usepackage{array,multirow,graphicx}
\usepackage{dcolumn}
\usepackage{newlfont}
\usepackage{bm}
\usepackage[colorlinks,citecolor=blue,urlcolor=blue,linkcolor=blue]{hyperref}
\usepackage[figtopcap]{subfigure}
\usepackage{color}

\usepackage{scalerel}
\usepackage{tikz}
\usetikzlibrary{svg.path}
\definecolor{orcidlogocol}{HTML}{A6CE39}
\tikzset{ 
  orcidlogo/.pic={
    \fill[orcidlogocol] svg{M256,128c0,70.7-57.3,128-128,128C57.3,256,0,198.7,0,128C0,57.3,57.3,0,128,0C198.7,0,256,57.3,256,128z};
    \fill[white] svg{M86.3,186.2H70.9V79.1h15.4v48.4V186.2z}
                 svg{M108.9,79.1h41.6c39.6,0,57,28.3,57,53.6c0,27.5-21.5,53.6-56.8,53.6h-41.8V79.1z M124.3,172.4h24.5c34.9,0,42.9-26.5,42.9-39.7c0-21.5-13.7-39.7-43.7-39.7h-23.7V172.4z}
                 svg{M88.7,56.8c0,5.5-4.5,10.1-10.1,10.1c-5.6,0-10.1-4.6-10.1-10.1c0-5.6,4.5-10.1,10.1-10.1C84.2,46.7,88.7,51.3,88.7,56.8z};}}
\newcommand\orcid[1]{\href{https://orcid.org/#1}{\mbox{\scalerel*{
\begin{tikzpicture}[yscale=-1,transform shape]
\pic{orcidlogo};
\end{tikzpicture}
}{|}}}}
\begin{document}
\title{A Minimal and Stable Vacuum Bounce in Exponential $f(R)$ Gravity}

\author{G.~G.~L.~Nashed$^{1,2}$~\orcid{0000-0001-5544-1119}}
\email{nashed@bue.edu.eg}
\author{A.~Eid$^{3}$}\email{amaid@imamu.edu.sa}
\affiliation {$^{1}$ Centre for Theoretical Physics, The British University, P.O. Box
43, El Sherouk City, Cairo 11837, Egypt\\$^{2}$
Centre for Space Research, North-West University, Potchefstroom, South Africa,
$^{4}$Department of Physics, College of Science, Imam Mohammad Ibn Saud Islamic University (IMSIU), Riyadh, Kingdom of Saudi Arabia}
\begin{abstract}
We investigate the realization of a nonsingular cosmological bounce in metric $f(R)$ gravity using a controlled exponential deformation of the Starobinsky $R^{2}$ model. Adopting a smooth Gaussian-type bouncing scale factor, we first demonstrate a no-go result showing that a positive-curvature vacuum bounce cannot be supported by the model
$f(R)=R+\alpha R^{2}(1-e^{-R/R_b})$ alone. We then show that a minimal extension obtained by introducing a constant term restores the bounce exactly, with the constant fixed algebraically by the bounce condition. A systematic parameter-space scan is performed to identify regions free of ghost and tachyonic instabilities. Working in the Einstein frame, we study the evolution of scalar and tensor perturbations across the bounce and show that both remain finite and well behaved. Our results establish a minimal, perturbatively stable realization of a vacuum bounce in $f(R)$ gravity that goes beyond background-level constructions.
\end{abstract}
\maketitle
\section{Introduction}
\label{sec:introduction}

Cosmological bouncing scenarios have long been explored as a promising alternative to the standard Big Bang paradigm, as they replace the initial singularity with a smooth transition from a contracting to an expanding phase. Early work on nonsingular cosmologies highlighted the possibility of avoiding initial singularities through modified gravitational dynamics or effective matter sources, while later studies developed systematic frameworks for bounce realizations in different theoretical settings \cite{Brandenberger:2016vhg,Novello:2008ra,Nashed:2021pkc,Brandenberger:2012zb}. Within General Relativity, (GR) however, the realization of a bounce typically requires violations of the null energy condition, often leading to instabilities or theoretical inconsistencies \cite{Rubakov:2014jja}. These difficulties have motivated extensive investigations of bouncing solutions in theories beyond GR, where purely geometric effects can play a central role \cite{Nojiri:2010wj,Capozziello:2011et}.

Modified theories of gravity provide a natural arena for constructing nonsingular cosmological models, since higher-curvature corrections or additional gravitational degrees of freedom can effectively generate repulsive gravitational effects at high curvature. Bouncing solutions have been investigated in a wide variety of modified gravity frameworks, including scalar--tensor theories, Gauss--Bonnet gravity, nonlocal gravity, and Horava--Lifshitz gravity \cite{Battefeld:2014uga,Cai:2014xxa,Cai:2013kja,Odintsov:2015gba}. In many of these models, the bounce arises without the need for exotic matter sources, highlighting the geometric origin of the nonsingular evolution.

Among modified gravity theories, metric $f(R)$ gravity represents one of the most conservative and well-studied extensions of GR~\cite{DeFelice:2010aj,Sotiriou:2008rp}. Its equivalence to a scalar--tensor theory with a single additional scalar degree of freedom makes it a natural framework for studying nonsingular early-universe dynamics~\cite{Whitt:1984pd,Maeda:1987xf}. Indeed, a variety of bouncing solutions have been proposed in $f(R)$ gravity using reconstruction techniques or specific functional choices~\cite{Nojiri:2010wj,Cai:2013kja,Cai:2012va}. However, many of these constructions focus primarily on background evolution and do not systematically address stability, perturbative behavior, or the minimality of the required modifications~\cite{Brandenberger:2016vhg}.

A number of works have constructed bouncing solutions within $f(R)$ gravity using reconstruction or designer techniques, in which the functional form of $f(R)$ is determined to reproduce a prescribed bouncing scale factor or Hubble evolution \cite{Cai:2014xxa,Cai:2013kja,Nashed:2023pxd,Elizalde:2014uba}. While such approaches are useful for generating explicit background solutions, they often obscure the minimality of the underlying gravitational modification and do not always allow for a transparent assessment of stability across the full curvature range. In particular, the behavior of scalar and tensor perturbations across the bounce has not been systematically addressed in many of these constructions \cite{Brandenberger:2016vhg,Nashed:2021gkp,Battefeld:2014uga}.

In this work, we adopt a different strategy. We begin with a simple and physically transparent bouncing background and examine whether a controlled exponential deformation of the Starobinsky $R^2$ model can support a vacuum bounce~\cite{Starobinsky:1980te}. We demonstrate that, despite its appealing high-curvature behavior, the pure exponential switch-on model fails to satisfy the vacuum bounce condition at positive curvature. This no-go result provides a clear diagnostic of the limitations of this class of models and is closely related to general consistency constraints in higher-curvature gravity~\cite{Brandenberger:2016vhg}. We then show that a minimal extension obtained by adding a constant term removes this obstruction without introducing new degrees of freedom or altering the stability structure of the theory. The constant is not treated as a free parameter but is fixed uniquely by the bounce condition. We perform a systematic parameter-space analysis to identify ghost-free and tachyon-free regions and study the evolution of scalar and tensor perturbations across the bounce in the Einstein frame~\cite{DeFelice:2010aj,Mukhanov:1990me,Sasaki:1986hm}. The resulting framework constitutes a clean and perturbatively stable realization of a vacuum bounce in $f(R)$ gravity, bridging the gap between purely background-level constructions and fully dynamical analyses~\cite{DeFelice:2010aj,Brandenberger:2016vhg}. Our approach provides a controlled benchmark for future studies of nonsingular cosmology in higher-curvature gravity~\cite{Novello:2008ra}.

It is important to emphasize how the present construction differs from earlier realizations of bouncing
cosmologies in $f(R)$ gravity. Many existing studies rely on reconstruction techniques or designer approaches,
in which the functional form of $f(R)$ is determined \emph{a posteriori} to reproduce a prescribed background
evolution~\cite{Nojiri:2010wj,Cai:2013kja,Cai:2012va}. While such methods are useful at the background level, they often
obscure the minimality of the required modifications and do not always allow for a transparent assessment of
stability across the full curvature history~\cite{DeFelice:2010aj,Brandenberger:2016vhg}.

In contrast, the approach adopted here starts from a simple and physically motivated class of
higher-curvature models and imposes the vacuum bounce condition exactly at the level of the modified
Friedmann equations~\cite{DeFelice:2010aj}. This strategy leads to a sharp and model-independent
no-go result: a positive-curvature vacuum bounce cannot be supported by the exponential switch-on
deformation of the Starobinsky model $f(R)=R+\alpha R^{2}(1-e^{-R/R_{b}})$ alone~\cite{Starobinsky:1980te}.
To our knowledge, this obstruction has not been identified previously in the literature.

Moreover, the resolution of this no-go result is achieved through a minimal extension involving a constant
term whose value is fixed algebraically by the bounce condition itself. This term does not introduce new
degrees of freedom, does not alter the high-curvature behavior responsible for the bounce, and preserves the
ghost- and tachyon-free structure of the theory~\cite{DeFelice:2010aj,Sotiriou:2008rp}. As a result,
the framework presented here provides a clean and controlled realization of a vacuum bounce that goes
beyond background-level constructions and allows for a fully consistent analysis of tensor and scalar
perturbations across the bounce~\cite{Mukhanov:1990me,Brandenberger:2016vhg}.


The paper is organized as follows. In Sec.~\ref{back}
 we introduce the bouncing background and derive the associated
geometric quantities. In Sec.~\ref{fR} we define the class of $f(R)$ models under consideration and discuss their
basic properties and stability conditions. In Sec.~\ref{NG} we impose the vacuum bounce condition and derive a
no-go result for the pure exponential switch-on model. In Sec.~\ref{RB} we show how this obstruction can be removed
by a minimal extension of the theory. In Sec.~\ref{SA} we perform a systematic parameter-space analysis and identify
viable regions. In Sec.~\ref{back} and Sec.~\ref{sec:perturbations} we study the Einstein-frame formulation and the evolution of tensor
and scalar perturbations across the bounce. In Sec.~\ref{sec:results} main numerical findings of the parameter-space scan and the perturbation analysis are presented.
 Our conclusions and outlook are presented in Sec.~\ref{sec:conclusions}.

\section{Background Bounce Ansatz and Geometric Quantities}\label{back}

In this work we consider a spatially flat FLRW universe with line element
\begin{equation}
ds^{2}=-dt^{2}+a^{2}(t)\,, \qquad
d\vec{x}^{\,2}\equiv dx^{2}+dy^{2}+dz^{2}.
\end{equation}
Adopting the symmetric Gaussian-type bouncing scale factor
\begin{equation}
a(t)=a_{b}\,e^{\lambda t^{2}}, \qquad a_b>0,\ \lambda>0.
\label{eq:scale_factor_bounce}
\end{equation}
This form ensures that the scale factor reaches a finite minimum at the bounce point $t=0$, where the
Hubble parameter vanishes, and evolves smoothly and symmetrically for $t>0$ and $t<0$. Gaussian and
exponential bouncing backgrounds of this type have been widely employed as analytically tractable and
maximally regular benchmarks in studies of nonsingular cosmology
\cite{Brandenberger:2016vhg,Novello:2008ra}.

The corresponding Hubble parameter is
\begin{equation}
H(t)\equiv \frac{\dot a}{a}=2\lambda t, \qquad \mbox{and its time derivative is} \qquad
\dot H(t)=2\lambda.
\label{eq:Hdot}
\end{equation}
Hence at the bounce time $t=t_b=0$ one has
\begin{equation}
H(0)=0, \qquad \dot H(0)=2\lambda>0,
\end{equation}
which confirms a nonsingular bounce (i.e. a transition from contraction to expansion).

For a flat FLRW background, the Ricci scalar is
\begin{equation}
R(t)=6\left(\dot H+2H^{2}\right).
\label{eq:Ricci_general}
\end{equation}
Substituting Eq.~\eqref{eq:Hdot} into Eq.~\eqref{eq:Ricci_general}, we obtain
\begin{align}
R(t)
&=6\left(2\lambda+2(2\lambda t)^{2}\right)
=12\lambda+48\lambda^{2}t^{2}.
\label{eq:Ricci_bounce}
\end{align}
Before specifying the functional form of $f(R)$, it is instructive to examine the
curvature evolution associated with the bouncing background.
\begin{figure}[t]
  \centering
  \subfigure[~Gaussian-type bouncing scale factor $a(t)$]{\label{fig:1}\includegraphics[width=0.25\textwidth]{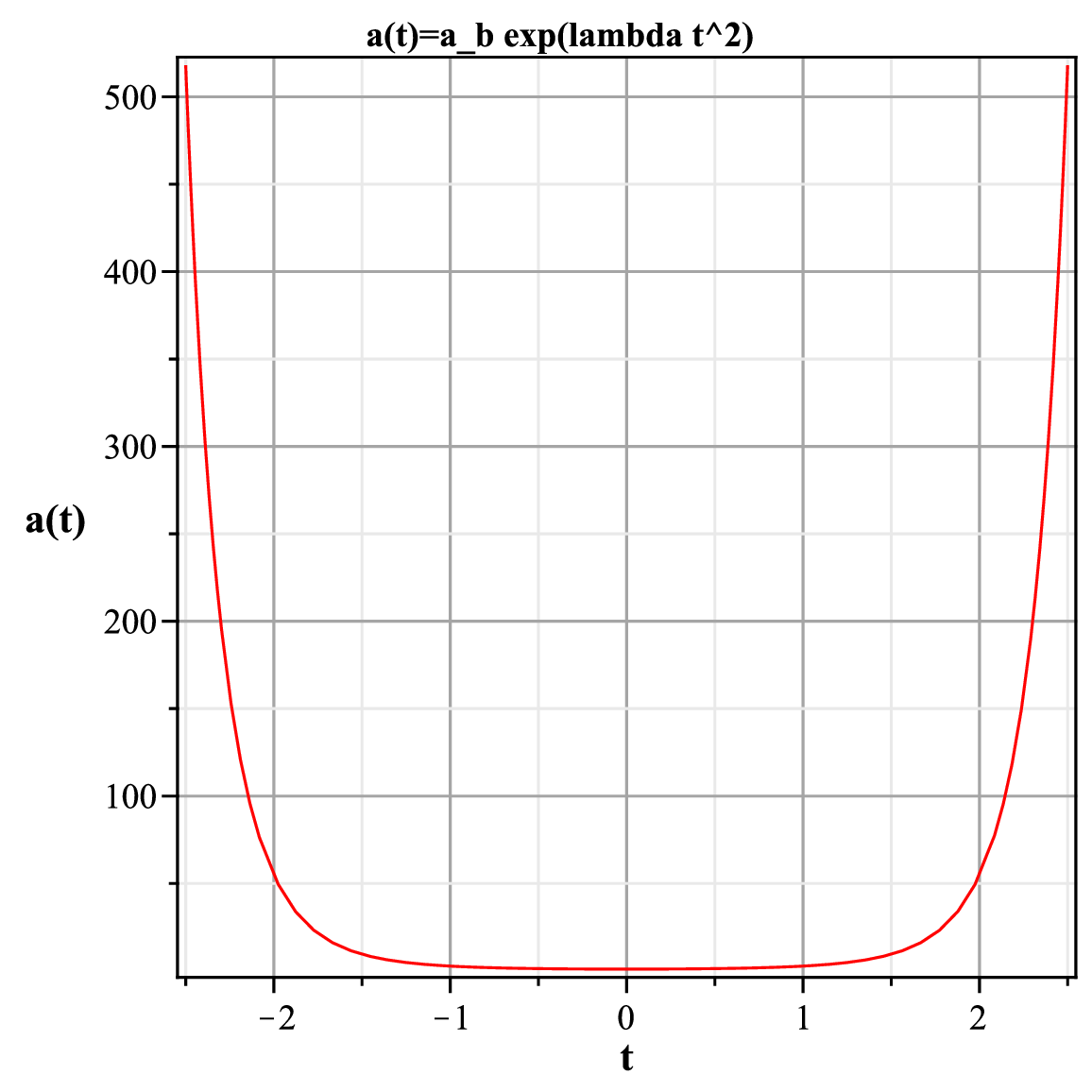}}
  \subfigure[~Phase portrait in the
$(a, \dot a)$ plane]{\label{fig:2}\includegraphics[width=0.25\textwidth]{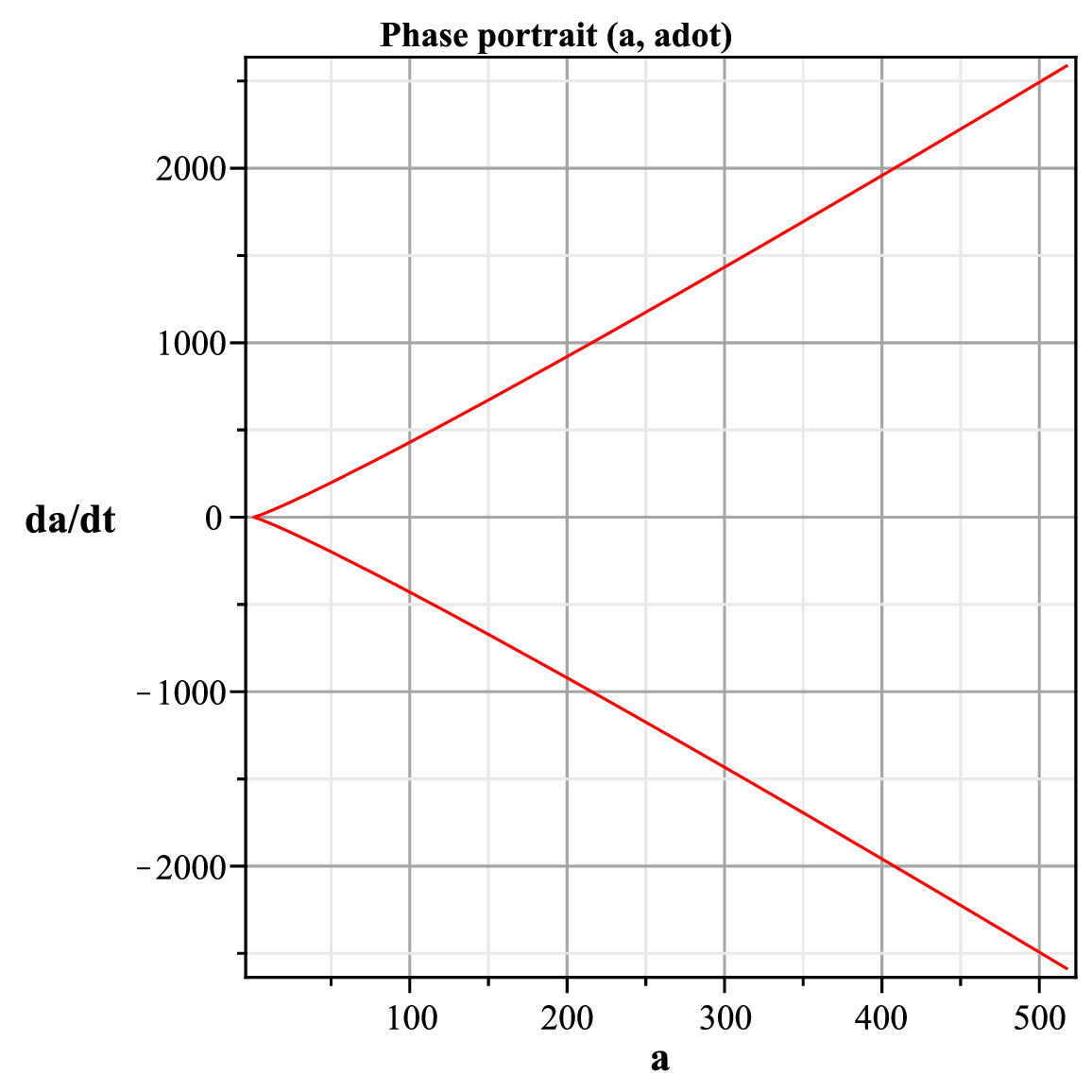}}
 \subfigure[~ Phase portrait in the $(H, \dot H)$ plane]{\label{fig:3} \includegraphics[width=0.25\textwidth]{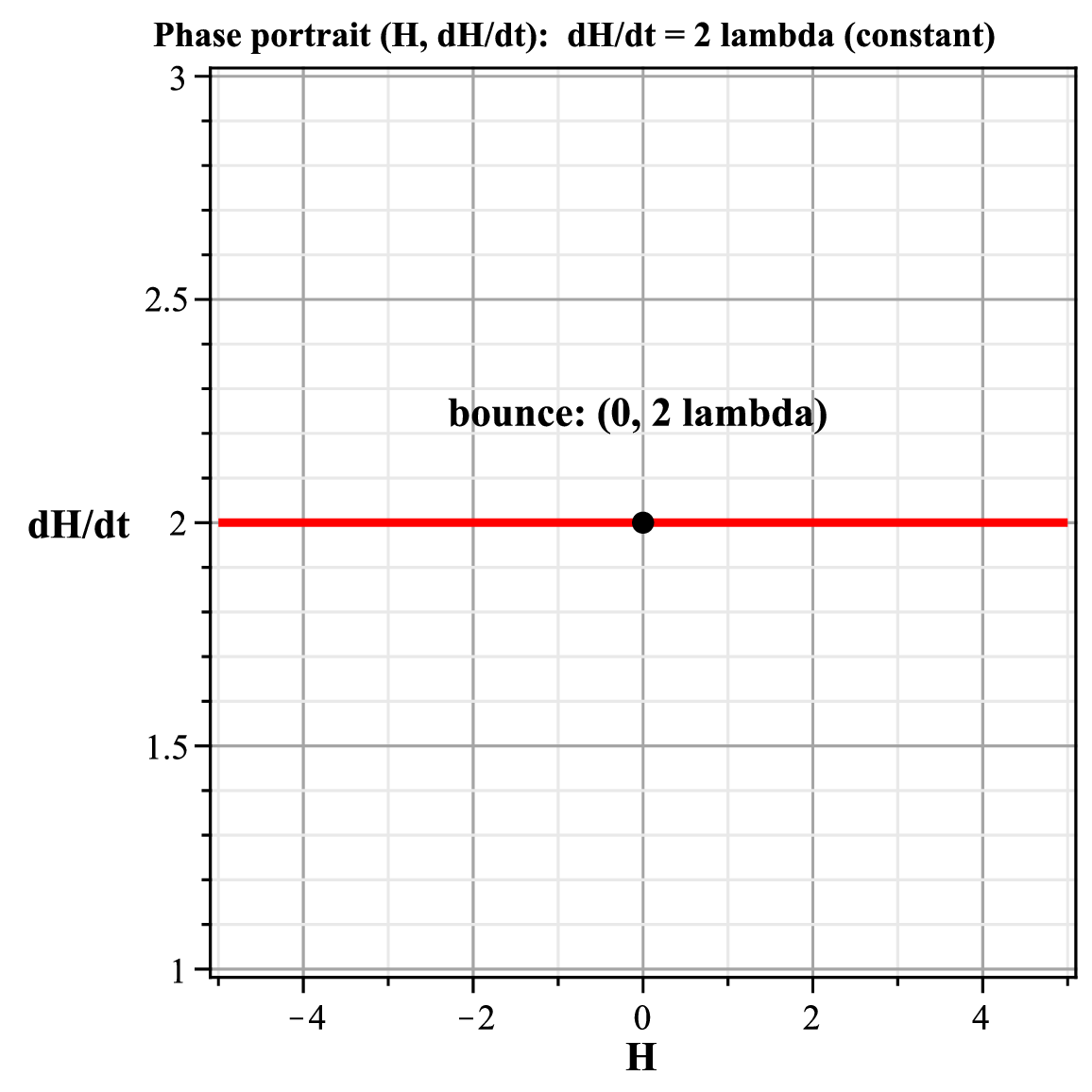}}
    \caption{\subref{fig:1} The Gaussian-type bouncing scale factor $a(t)=a_b e^{\lambda t^2}$,
    \subref{fig:2} Phase portrait in the $(a,\dot a)$ plane, showing that the trajectory passes smoothly through $\dot a=0$,
confirming the nonsingular bounce; \subref{fig:3} Phase portrait in the $(H,\dot H)$ plane. The straight-line trajectory reflects the fact that $\dot H=2\lambda$
is constant, corresponding to a maximally regular bounce with positive acceleration at $H=0$.
}
  \label{fig:phas}
\end{figure}

\begin{figure}[t]
  \centering
  \subfigure[~Time evolution of the Ricci scalar $R(t)$]{\label{fig:7}\includegraphics[width=0.25\textwidth]{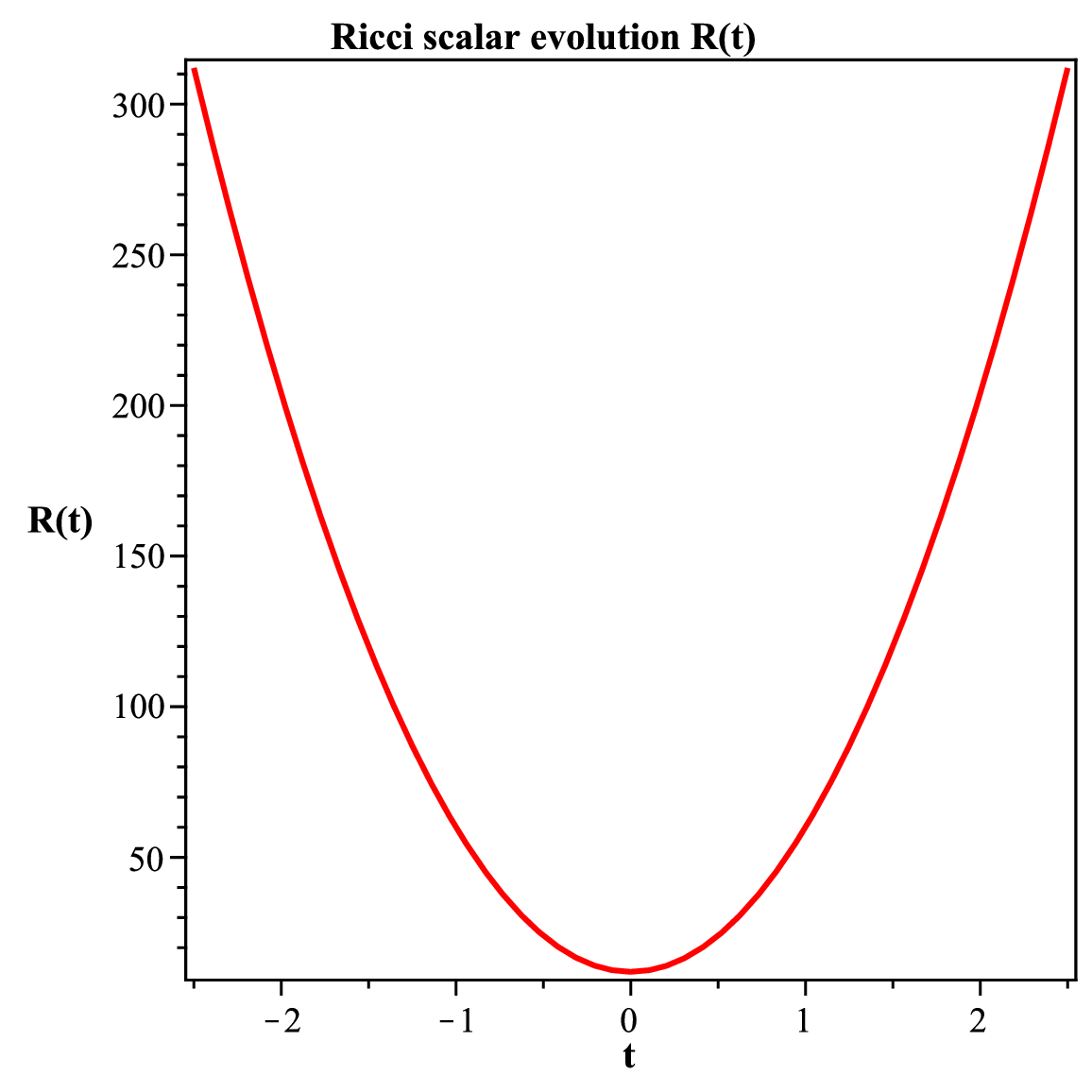}}
  \subfigure[~Phase portrait in the $(R,\dot R)$ plane]{\label{fig:8}\includegraphics[width=0.25\textwidth]{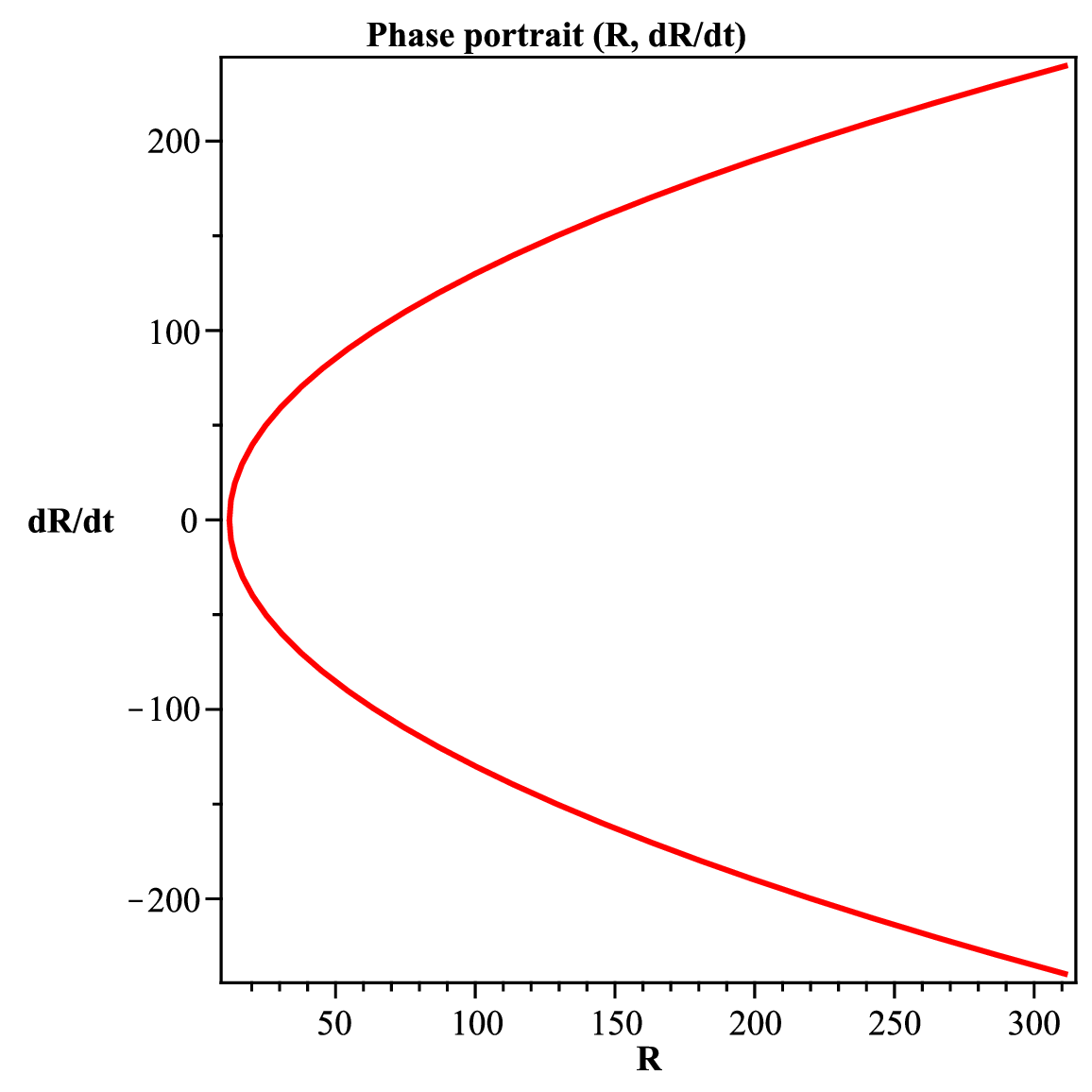}}
   \caption{Curvature evolution associated with the Gaussian bouncing background
$a(t)=a_b e^{\lambda t^2}$.
Panel (a) shows the time dependence of the Ricci scalar
$R(t)=12\lambda+48\lambda^2 t^2$, which remains finite and positive throughout
the evolution and attains its minimum value at the bounce point $t=0$.
Panel (b) displays the phase portrait in the $(R,\dot R)$ plane, where
$\dot R(t)=96\lambda^2 t$.
The trajectory crosses $\dot R=0$ smoothly at finite $R$, indicating that the
bounce is free of curvature singularities and that the curvature dynamics are
regular across the transition from contraction to expansion.
}
\label{fig:ricci_phase}
\end{figure}
The curvature history displayed in Fig.~\ref{fig:ricci_phase}~\subref{fig:7} determines the range of $R$ over which the
stability conditions of the $f(R)$ theory must be satisfied \cite{Brandenberger:2016vhg}.


Therefore the curvature at the bounce is finite and given by
\begin{equation}
R_{0}\equiv R(0)=12\lambda, \quad \mbox {and} \qquad  R(t)\ge R_{0} \qquad \mbox{for all t}.
\label{eq:R0}
\end{equation}
The behaviors of the scale factor (\ref{eq:scale_factor_bounce}) and its derivative are given in Fig.~\ref{fig:phas}~\subref{fig:1}. Figure~\ref{fig:phas}~\subref{fig:2} shows the phase portrait in the $(a,\dot a)$ plane corresponding to the Gaussian bounce. The trajectory smoothly reaches $\dot a=0$ at a finite, nonzero value of the scale factor,
indicating that the bounce occurs without any divergence in the expansion rate.
Figure~\ref{fig:phas}~\subref{fig:3} displays the phase portrait in the $(H,\dot H)$ plane. In this case,
the trajectory reduces to a straight line, reflecting the fact that the time derivative of the Hubble parameter
is constant, $\dot H=2\lambda$. The bounce point corresponds to $(H,\dot H)=(0,2\lambda)$, where the Hubble parameter
changes sign while the acceleration remains positive. This straight-line behavior indicates that the bounce is
maximally regular and symmetric, with a constant effective acceleration driving the transition between contraction
and expansion.

The Gaussian-type bouncing scale factor adopted in this work is not chosen merely for mathematical
convenience, but rather for its maximally regular and physically transparent properties.
The background is fully symmetric around the bounce point and is characterized by a constant and
positive time derivative of the Hubble parameter, $\dot H = 2\lambda$, which ensures a smooth and
controlled transition between contraction and expansion.

This choice is particularly advantageous for testing higher-curvature gravity models.
First, all relevant geometric quantities, including $a(t)$, $H(t)$, $\dot H(t)$, and $R(t)$, remain finite
and analytic throughout the evolution, eliminating any ambiguity associated with coordinate or
curvature singularities. Second, the monotonic and strictly positive behavior of the Ricci scalar
provides a well-defined curvature history over which the viability and stability conditions of the
$f(R)$ theory can be imposed consistently.


\section{\texorpdfstring{$f(R)$}{f(R)} Model Family and Basic Derivatives}\label{fR}

We work in the metric formulation of $f(R)$ gravity, described by the action
\begin{equation}
S=\frac{1}{2\kappa^{2}}\int d^{4}x\,\sqrt{-g}\,f(R)\;+\;S_{m},
\label{eq:action_fR}
\end{equation}
where $\kappa^{2}=8\pi G$, $g$ is the determinant of the metric $g_{\mu\nu}$, $R$ is the Ricci scalar, and $S_m$ denotes the matter action. In what follows, we focus on the vacuum case unless stated otherwise, i.e. $S_m=0$.

\subsection{Choice of the model}

To generate a nonsingular bounce from purely geometric effects, we consider an ``exponential switch-on'' deformation of the Starobinsky-type $R^{2}$ correction,
\begin{equation}
f(R)=R+\alpha R^{2}\left(1-e^{-R/R_{b}}\right),
\label{eq:fR_model}
\end{equation}
where $\alpha$ and $R_b$ are free parameters. The exponential factor ensures that the $R^2$ correction becomes relevant at sufficiently large curvature (e.g. near the bounce), while it is suppressed at low curvature. This provides a controlled modification of GR  away from the high-curvature regime.

It is useful to examine two limits of the model~\eqref{eq:fR_model}.

\paragraph{Low-curvature regime ($R\ll R_b$).}
Expanding the exponential,
\begin{align}
e^{-R/R_b}&=1-\frac{R}{R_b}+\frac{R^2}{2R_b^2}+\mathcal{O}\!\left(\frac{R^3}{R_b^3}\right), \quad \Rightarrow
f(R)
=R+\alpha\left(\frac{R^3}{R_b}-\frac{R^4}{2R_b^2}+\cdots\right).
\label{eq:lowR_expansion}
\end{align}
Thus the leading correction at small curvature is $\sim R^3/R_b$, i.e. the deviation from GR is strongly suppressed.

\paragraph{High-curvature regime ($R\gg R_b$).}
When $R/R_b\to\infty$, the exponential term vanishes, $e^{-R/R_b}\to 0$, and therefore
\begin{equation}
f(R)\;\longrightarrow\;R+\alpha R^2,
\label{eq:highR_limit}
\end{equation}
which is the well-known Starobinsky form. Hence near sufficiently large curvature the model effectively behaves as an $R+\alpha R^2$ theory.

The modified field equations and stability conditions depend on the first and second derivatives of $f(R)$ with respect to $R$. We define
\begin{equation}
f_{R}(R)\equiv \frac{df}{dR}, \qquad f_{RR}(R)\equiv \frac{d^{2}f}{dR^{2}}. \quad \mbox{For compactness,  we introduce} \quad
E(R)\equiv e^{-R/R_b}.
\label{eq:E_def}
\end{equation}
Differentiating Eq.~\eqref{eq:fR_model}, we obtain
\begin{align}
f_{R}(R)
&=1+\alpha\frac{d}{dR}\left[R^{2}\left(1-E(R)\right)\right] =1+\alpha\left[2R\left(1-e^{-R/R_b}\right)+\frac{R^{2}}{R_b}e^{-R/R_b}\right].
\label{eq:fR_boxed}
\end{align}

Differentiating once more, we find
\begin{align}
f_{RR}(R)
&=\alpha\frac{d}{dR}\left[2R(1-E)+\frac{R^{2}}{R_b}E\right]=\alpha\left[2\left(1-e^{-R/R_b}\right)+\frac{4R}{R_b}e^{-R/R_b}-\frac{R^{2}}{R_b^{2}}e^{-R/R_b}\right].
\label{eq:fRR_boxed}
\end{align}
In metric $f(R)$ gravity, the absence of ghosts and tachyonic instabilities is commonly ensured (at least at the background level) by the conditions
\begin{equation}
f_R(R)>0,
\qquad
f_{RR}(R)>0.
\label{eq:viability_conditions}
\end{equation}
The first inequality guarantees an effective positive gravitational coupling, while the second inequality ensures that the extra scalar degree of freedom (the scalaron) is not tachyonic in the regime of interest. In the present work, these conditions will be imposed throughout a time interval around the bounce, $t\in[-T,T]$, where $T$ is chosen such that $R(t)$ spans the relevant curvature range for the bouncing phase.

\section{Vacuum Bounce Condition and a No-Go Result}\label{NG}

In this section we impose the bounce condition directly on the modified Friedmann equations of metric $f(R)$ gravity and derive a key result for the model introduced in the previous section.
Varying the action~\eqref{eq:action_fR} with respect to the metric yields the field equations
\begin{equation}
f_R R_{\mu\nu}-\frac{1}{2}f g_{\mu\nu}
-\nabla_\mu\nabla_\nu f_R
+g_{\mu\nu}\Box f_R
=\kappa^2 T_{\mu\nu}.
\end{equation}
For a spatially flat FLRW background, the $(00)$ component can be written as
\begin{equation}
3H^2 f_R
=\frac{1}{2}\left(f_R R-f\right)
-3H\dot f_R
+\kappa^2\rho ,
\label{eq:Friedmann_fR}
\end{equation}
where $\rho$ is the energy density of matter. Throughout this section we assume a \emph{vacuum bounce}, i.e.
\begin{equation}
\rho=0.
\end{equation}

At the bounce time $t=t_b=0$, the Hubble parameter vanishes,
\begin{equation}
H(0)=0,
\end{equation}
while the Ricci scalar takes the finite value $R_0=12\lambda$ (see Eq.~\eqref{eq:R0}). Evaluating Eq.~\eqref{eq:Friedmann_fR} at the bounce, all terms proportional to $H$ vanish, and we obtain the algebraic constraint
\begin{equation}
0=\frac{1}{2}\left[f_R(R_0)\,R_0-f(R_0)\right].
\end{equation}
Equivalently, the vacuum bounce condition can be written as
\begin{equation}
{
f(R_0)=f_R(R_0)\,R_0 .
}
\label{eq:bounce_condition}
\end{equation}
This condition is purely geometric and must be satisfied by any $f(R)$ model that supports a vacuum bounce with $R_0\neq 0$.

We now apply Eq.~\eqref{eq:bounce_condition} to the specific model
\begin{equation}\label{fR}
f(R)=R+\alpha R^2\left(1-e^{-R/R_b}\right).
\end{equation}
Using the expressions derived in Eqs.~\eqref{eq:fR_boxed} and~\eqref{eq:fR_model}, we compute
\begin{align}
f_R(R)\,R - f(R)
&=\alpha\Bigg[
R^2\left(1-e^{-R/R_b}\right)
+\frac{R^3}{R_b}e^{-R/R_b}
\Bigg]=\alpha\left[
R_0^2\left(1-e^{-R_0/R_b}\right)
+\frac{R_0^3}{R_b}e^{-R_0/R_b}
\right]_{\mid_{R=R_0}}.
\label{eq:bounce_expression}
\end{align}

For a genuine bounce of the type considered here, one has
\begin{equation}
R_0=12\lambda>0,
\qquad
R_b>0.
\end{equation}
Under these assumptions,
\begin{itemize}
\item $1-e^{-R_0/R_b} > 0$,
\item $e^{-R_0/R_b} > 0$,
\item $R_0^2>0$ and $R_0^3/R_b>0$.
\end{itemize}
Therefore, the quantity inside the brackets in Eq.~\eqref{eq:bounce_expression} is strictly positive. As a consequence,
\begin{equation}
f_R(R_0)\,R_0 - f(R_0) > 0
\qquad \text{for } \alpha>0, \quad \mbox{and} \quad
f_R(R_0)\,R_0 - f(R_0) < 0
\qquad \text{for } \alpha<0.
\end{equation}
In either case, the bounce condition~\eqref{eq:bounce_condition} can only be satisfied if $\alpha=0$ which corresponds to GR, for which the higher-curvature modification vanishes and the bounce cannot be supported by geometry alone. We thus arrive at the following conclusion:
\begin{quote}
\emph{For the exponential switch-on model
$f(R)=R+\alpha R^2(1-e^{-R/R_b})$ with $R_0>0$ and $R_b>0$, a vacuum cosmological bounce satisfying $H(0)=0$ and $R_0\neq 0$ is not possible unless the modification parameter $\alpha$ vanishes.}
\end{quote}

Although the no-go result derived above was obtained using the Gaussian bouncing background,
its origin is more general and does not rely on the detailed time dependence of the scale factor.
The obstruction arises from the purely algebraic structure of the vacuum bounce condition
$f(R_0)=R_0 f_R(R_0)$ evaluated at a positive and finite curvature $R_0>0$.
For the exponential switch-on model, the combination $R f_R - f$ is sign-definite at positive
curvature, independently of the specific form of $R(t)$ away from the bounce.

As a consequence, any smooth vacuum bounce characterized by $H(t_b)=0$ and $R(t_b)>0$ will face
the same obstruction within this class of models, provided the curvature remains finite and
positive at the bounce. The no-go result therefore reflects a structural limitation of the
exponential switch-on deformation of the Starobinsky model, rather than a peculiarity of the
chosen background ansatz.

This observation highlights the significance of the result: the failure of the pure model to
support a positive-curvature vacuum bounce is robust and generic within its domain of validity,
thereby providing a clear and well-defined motivation for the minimal extension introduced
in the next section.

No-go results of this type are particularly valuable, as they identify genuine structural
limitations of modified gravity models and help delineate which extensions are necessary for
constructing consistent nonsingular cosmologies.

This constitutes a no-go result for a positive-curvature vacuum bounce in the present model. In the next section, we will show how this obstruction can be removed by introducing a minimal extension of the theory, while preserving the desirable features of the exponential switch-on mechanism.

\section{Minimal Extension: Restoring the Bounce with a Constant Term}\label{RB}

The no-go result obtained in the previous section shows that the exponential switch-on model
\(
f(R)=R+\alpha R^2(1-e^{-R/R_b})
\)
cannot support a positive-curvature vacuum bounce by itself. In this section, we demonstrate that a \emph{minimal and well-motivated extension} of the model is sufficient to restore an exact vacuum bounce, while preserving the controlled high-curvature behavior.


We consider the following extension of the gravitational Lagrangian:
\begin{equation}
f(R)=R-2\Lambda+\alpha R^2\left(1-e^{-R/R_b}\right),
\label{eq:fR_extended}
\end{equation}
where $\Lambda$ is a constant parameter. This term plays a role analogous to a cosmological constant, but here it is not introduced to drive late-time acceleration; instead, it is fixed by the bounce condition itself.

Importantly, the addition of $-2\Lambda$ does not alter the functional form of $f_R$ and $f_{RR}$
so that the ghost- and tachyon-free conditions remain unchanged.


For the extended model~\eqref{eq:fR_extended}, we find
\begin{align}
f_R(R)\,R-f(R)
&=2\Lambda
+\alpha\left[
R^2\left(1-e^{-R/R_b}\right)
+\frac{R^3}{R_b}e^{-R/R_b}
\right].
\label{eq:fR_minus_f_ext}
\end{align}
Imposing Eq.~\eqref{eq:bounce_condition} at $R=R_0$ therefore yields
\begin{equation}
2\Lambda
=-\alpha\left[
R_0^2\left(1-e^{-R_0/R_b}\right)
+\frac{R_0^3}{R_b}e^{-R_0/R_b}
\right].
\label{eq:Lambda_solution}
\end{equation}


Equation~\eqref{eq:Lambda_solution} shows that, for any chosen values of $(\alpha,R_b)$, the constant $\Lambda$ can be fixed uniquely so that the vacuum bounce condition is satisfied exactly. In particular, for $R_0>0$ and $R_b>0$,
\begin{itemize}
\item if $\alpha>0$, then $\Lambda<0$,
\item if $\alpha<0$, then $\Lambda>0$.
\end{itemize}
Thus, the obstruction encountered in the pure model is removed by the presence of the constant term, and a nonsingular vacuum bounce becomes possible without introducing any matter component.

The role of $\Lambda$ in Eq.~\eqref{eq:fR_extended} is to compensate the strictly positive geometric contribution generated by the exponential $R^2$ term at the bounce curvature $R_0$. Since $\Lambda$ is fixed algebraically by the bounce condition, it does not represent an additional free parameter at this stage of the construction.

{
The constant term $2\Lambda$ introduced in Eq.~(\ref{eq:Lambda_solution}) should not be interpreted as a freely tunable cosmological constant in the usual phenomenological sense. Its value is fixed uniquely and algebraically by the vacuum bounce condition at the curvature scale $R_0$. Thus, $\Lambda$ is not adjusted to fit observational data nor introduced to drive late-time acceleration, but rather emerges as a geometric counterterm required to satisfy the modified Friedmann equation at $H=0$. At the bounce curvature $R_0>0$, the exponential $R^2$ sector contributes a sign-definite term to the combination $Rf_R - f$. The constant $-2\Lambda$ compensates this contribution, restoring algebraic consistency of the vacuum bounce condition without introducing new degrees of freedom or modifying the derivatives $f_R$ and $f_{RR}$. In curvature regimes far from the bounce, the relative importance of the constant term depends on the value of $R$. At high curvature ($R \gg R_b$), the $R^2$ correction dominates and the constant becomes subleading.
At sufficiently low curvature, the term may act effectively as a cosmological constant.
However, since its magnitude is fixed by the bounce condition rather than by late-time cosmological requirements, its interpretation differs conceptually from that of a standard vacuum energy parameter.
}

We emphasize that this extension is minimal in the sense that:
\begin{enumerate}
\item it does not modify the high-curvature behavior responsible for the bounce,
\item it leaves the stability conditions $f_R>0$ and $f_{RR}>0$ unchanged,
\item it introduces no new dynamical degree of freedom.
\end{enumerate}

The constant term $2\Lambda$ introduced in Eq.~(\ref{eq:Lambda_solution}) 
should not be interpreted as a freely tunable cosmological constant. 
Its value is fixed uniquely by the vacuum bounce condition at $R_0$.

As a result, this term does not introduce new dynamical degrees of freedom 
and does not modify the high-curvature behavior governed by the exponential $R^2$ term.

{
For clarity, we summarize the precise domain in which the above no-go result holds.
The obstruction applies strictly to metric $f(R)$ gravity in vacuum ($\rho=0$), for smooth nonsingular bounces characterized by finite and positive curvature at the bounce point ($R_0>0$), and for the specific exponential switch-on model given by Eq.~\eqref{fR} without an additional constant term, i.e., with $\Lambda=0$. The derivation relies on the algebraic bounce condition $f(R_0) = R_0 f_R(R_0)$
which follows directly from the modified Friedmann equation evaluated at $H=0$ in vacuum. Since the combination $R f_R - f$ is sign-definite at positive finite curvature for the model under consideration, the condition cannot be satisfied unless $\alpha=0$. The result does not exclude matter-bounce realizations, scenarios with $R_0=0$, negative-curvature bounces, Palatini $f(R)$ gravity, or more general higher-curvature extensions. It therefore reflects a structural limitation of the pure exponential switch-on model within its specific domain of validity.
}

In the next section, we will perform a systematic parameter scan in the $(\alpha,R_b)$ plane, determine the corresponding value of $\Lambda$ from Eq.~\eqref{eq:Lambda_solution}, and identify the regions in which the bounce is both viable and free of ghost and tachyonic instabilities.

\section{Parameter Space Analysis and Viability Conditions}\label{SA}

In this section we set up a systematic scan of the parameter space of the extended model~\eqref{eq:fR_extended} and formulate the viability conditions in a form suitable for numerical implementation.


To simplify the analysis and reduce the number of independent scales, we introduce dimensionless quantities normalized by the curvature at the bounce,
\begin{equation}
R_0 \equiv 12\lambda\quad \mbox{and we define} \quad
\bar R_b \equiv \frac{R_b}{R_0},
\qquad
\bar \alpha \equiv \alpha R_0,
\qquad
\bar R \equiv \frac{R}{R_0}.
\label{eq:dimensionless_defs}
\end{equation}
In terms of these variables, the exponential factor becomes
\begin{equation}
e^{-R/R_b}=e^{-\bar R/\bar R_b}.
\end{equation}


Using Eq.~\eqref{eq:Lambda_solution}, the constant $\Lambda$ required to enforce the vacuum bounce can be written in dimensionless form as
\begin{equation}
\Lambda
=-\frac{R_0}{2}\,\bar\alpha
\left[
\left(1-e^{-1/\bar R_b}\right)
+\frac{1}{\bar R_b}e^{-1/\bar R_b}
\right].
\label{eq:Lambda_dimensionless}
\end{equation}
Thus, once $(\bar\alpha,\bar R_b)$ are specified, $\Lambda$ is fixed uniquely and does not introduce an additional free parameter.


From Eq.~\eqref{eq:Ricci_bounce}, the Ricci scalar can be expressed in dimensionless form as
\begin{equation}
\bar R(t)=1+4\lambda t^{2}.
\label{eq:Rbar_t}
\end{equation}
In the numerical analysis, we restrict attention to a finite time interval
\begin{equation}
t\in[-T,T],
\end{equation}
where $T$ is chosen such that $\bar R(t)$ covers the curvature range relevant for the bouncing phase (typically $T\sim \mathcal{O}(1/\sqrt{\lambda})$).


Using Eqs.~\eqref{eq:fR_boxed} and~\eqref{eq:fRR_boxed}, the first derivative of $f(R)$ can be written as
\begin{align}
&f_R(\bar R)
=1+\bar\alpha\left[
2\bar R\left(1-e^{-\bar R/\bar R_b}\right)
+\frac{\bar R^{2}}{\bar R_b}e^{-\bar R/\bar R_b}
\right], \quad \mbox{while the second derivative becomes} \nonumber\\
&f_{RR}(\bar R)
=\frac{\bar\alpha}{R_0}\left[
2\left(1-e^{-\bar R/\bar R_b}\right)
+\frac{4\bar R}{\bar R_b}e^{-\bar R/\bar R_b}
-\frac{\bar R^{2}}{\bar R_b^{2}}e^{-\bar R/\bar R_b}
\right].
\label{eq:fRR_dimensionless}
\end{align}
Since $R_0>0$, the sign of $f_{RR}$ is controlled by the bracketed expression and the sign of $\bar\alpha$.

For each point in the parameter space $(\bar\alpha,\bar R_b)$, we put the following conditions throughout the interval $t\in[-T,T]$:
\begin{enumerate}
\item \textbf{No ghost condition:}
\begin{equation}
f_R(\bar R(t))>0.
\label{eq:ghost_condition}
\end{equation}

\item \textbf{No tachyonic instability:}
\begin{equation}
f_{RR}(\bar R(t))>0.
\label{eq:tachyon_condition}
\end{equation}
This ensures that the scalar degree of freedom associated with $f(R)$ gravity has a positive kinetic term and does not suffer from tachyonic instabilities near the bounce.

\item \textbf{Regularity at the bounce:}
\begin{equation}
f_R(\bar R=1)<\infty,
\qquad
f_{RR}(\bar R=1)<\infty.
\end{equation}
\end{enumerate}

\begin{figure}[t]
  \centering
  \subfigure[~ Viability map in the $(\bar{\alpha},\bar{R}_b)$ plane (classification $0/0.5/1$).]{\label{fig:4}\includegraphics[width=0.35\textwidth]{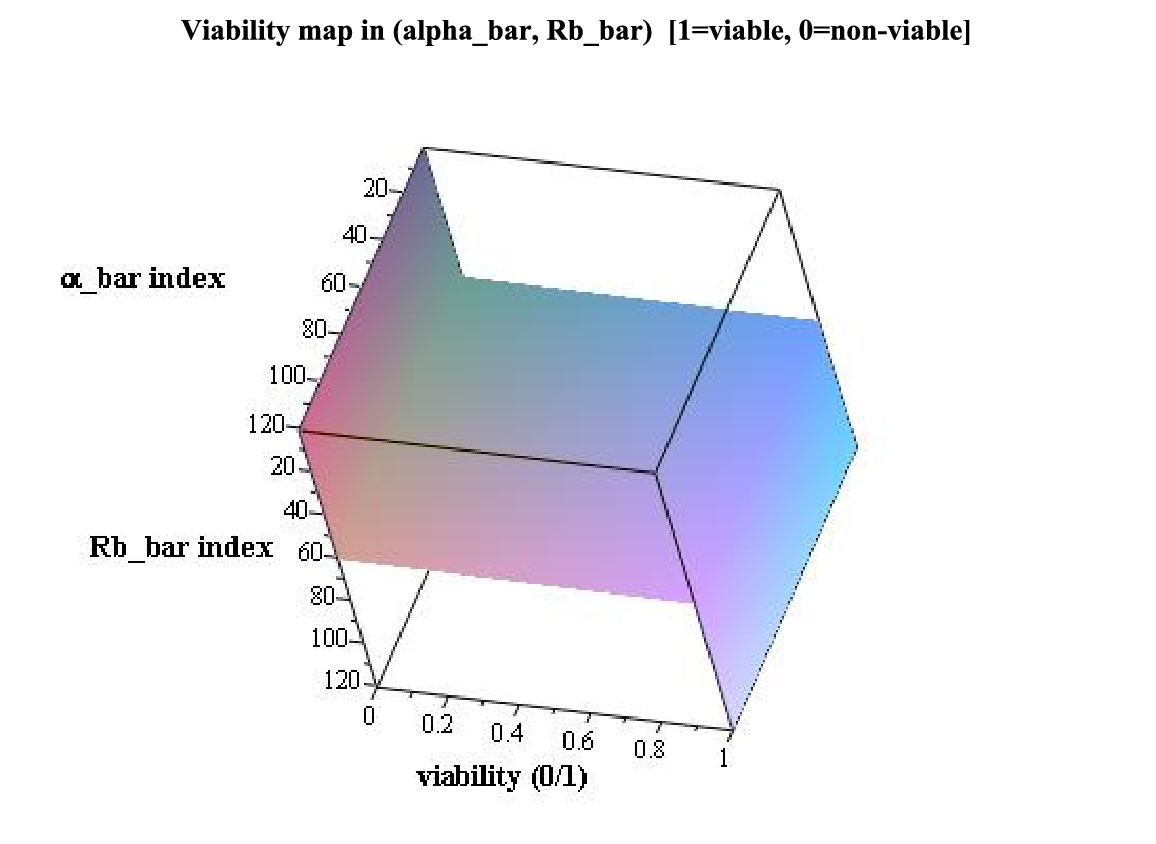}}
  \subfigure[~ Alternative view of the viability map (same data; top-view / compressed perspective).]{\label{fig:5}\includegraphics[width=0.35\textwidth]{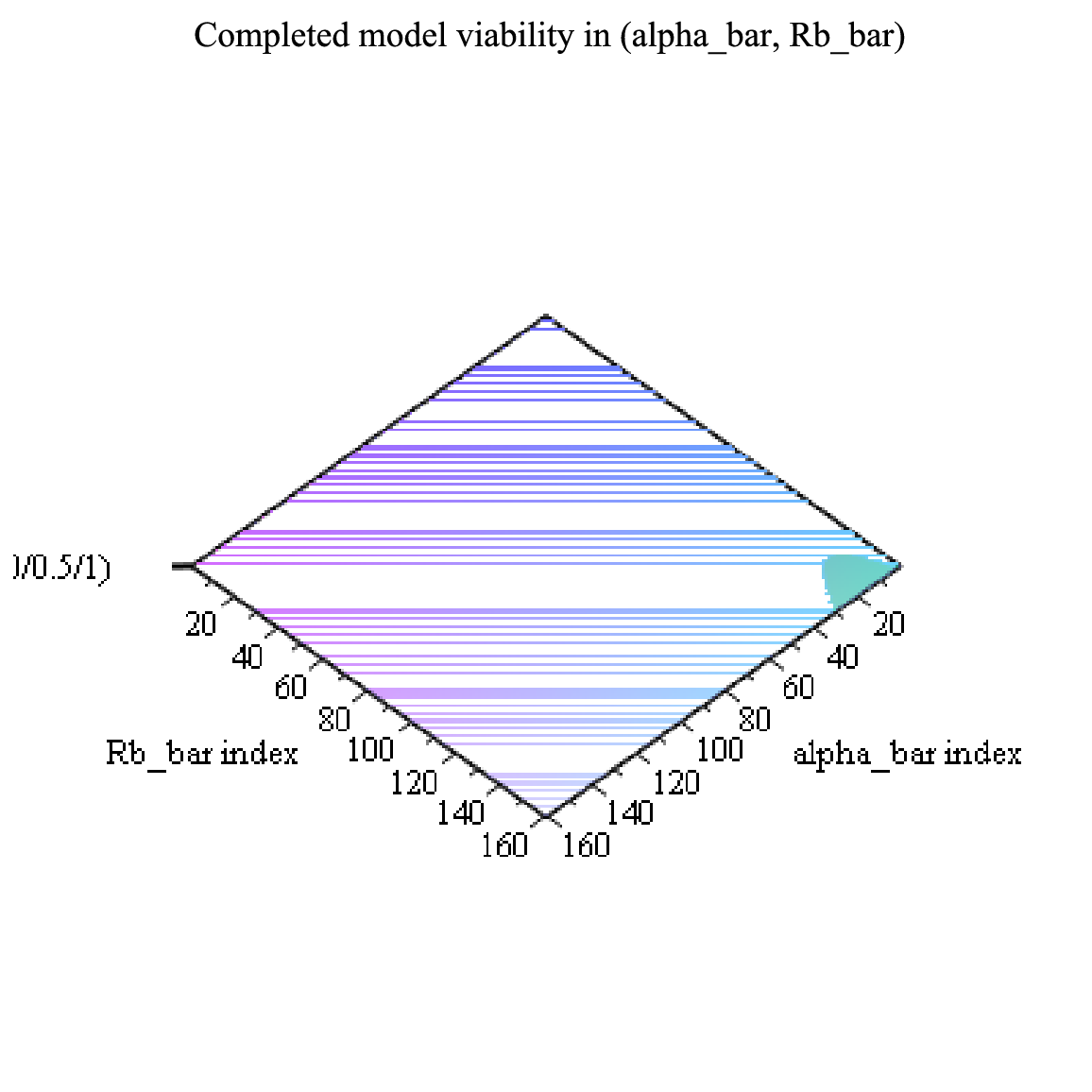}}
   \caption{
Parameter-space scan for the extended model
$f(R)=R-2\Lambda+\alpha R^2\!\left(1-e^{-R/R_b}\right)$
in terms of the dimensionless parameters $(\bar{\alpha},\bar{R}_b)$ defined in Eq.~(\ref{eq:dimensionless_defs}).
For each grid point, $\Lambda$ is fixed by the bounce condition via Eq.~(\ref{eq:fR_extended}), and the ghost- and tachyon-free
requirements $f_R(\bar{R}(t))>0$ and $f_{RR}(\bar{R}(t))>0$ are imposed throughout the bounce interval
$t\in[-T,T]$ as in Eqs.~(\ref{eq:ghost_condition})--(\ref{eq:tachyon_condition}), with $\bar{R}(t)=1+4\lambda t^2$ Eq.~(\ref{eq:Rbar_t}).
Panel \subref{fig:4} displays the resulting viability classification over the scan grid, while Panel \subref{fig:5} presents the same outcome
in an alternative view for readability.
}
\label{fig:region}
\end{figure}
Figure~\ref{fig:region} summarizes the outcome of the parameter-space scan in the
$(\bar{\alpha},\bar{R}_b)$ plane. The figure shows that viable vacuum bouncing solutions
exist only in restricted regions where both stability conditions $f_R>0$ and $f_{RR}>0$
are satisfied throughout the bouncing phase.
For moderate values of $\bar{R}_b=\mathcal{O}(1)$, an extended range of positive $\bar{\alpha}$
admits a stable bounce, indicating that the exponential $R^2$ correction can consistently
support the required geometric dynamics.
In contrast, very small values of $\bar{R}_b$ lead to rapid variations of $f_R$ and $f_{RR}$,
which typically violate the tachyon-free condition near the bounce.
For sufficiently large $\bar{R}_b$, the exponential term becomes suppressed and the model
approaches the pure $R+\alpha R^2$ limit, causing the viable region to shrink.
Overall, Fig.~\ref{fig:region} demonstrates that the vacuum bounce is not generic
but arises only for specific combinations of $(\bar{\alpha},\bar{R}_b)$, highlighting the
importance of the parameter scan in identifying physically consistent models.

For moderate values $\bar R_b=\mathcal{O}(1)$, the exponential transition is sufficiently smooth
to regulate the curvature dynamics without inducing instabilities. In this case, a continuous
range of positive $\bar\alpha$ supports a viable vacuum bounce, indicating that the exponential
$R^2$ sector can consistently provide the required geometric effects while preserving the stability
conditions throughout the bouncing phase.

In contrast, for large $\bar R_b$ the exponential factor becomes strongly suppressed over the
relevant curvature range, and the model approaches the pure $R+\alpha R^2$ limit.
As a result, the parameter space admitting a stable bounce shrinks, reflecting the fact that
the exponential deformation becomes dynamically ineffective.
These features explain why viable vacuum bounces arise only in restricted regions of the
$(\bar\alpha,\bar R_b)$ plane and highlight the importance of the parameter-space scan in identifying
physically consistent models \cite{Nashed:2009hn}.

The numerical scan described above provides the basis for selecting representative viable models,
which are analyzed in detail through their Einstein-frame perturbation dynamics in the next section.

\section{Einstein-Frame Formulation and Background Quantities for Perturbations}\label{EF}
\label{sec:einstein_frame}

The Einstein-frame formulation is adopted not only for conceptual clarity, but also because it
provides numerically stable and canonical evolution equations for cosmological perturbations.

In this section we construct the Einstein-frame representation of the viable Jordan-frame models identified in the parameter scan. This formulation is convenient for the numerical evolution of cosmological perturbations, since metric $f(R)$ gravity is dynamically equivalent to GR plus a canonical scalar field (the \emph{scalaron}) whenever $f_R>0$.


Assuming $f_R(R)>0$, we perform the conformal transformation
\begin{equation}
\tilde g_{\mu\nu} = f_R(R)\, g_{\mu\nu},
\label{eq:conformal_transform}
\end{equation}
where tilded quantities refer to the Einstein frame. Introducing the reduced Planck mass $M_{\textrm Pl}\equiv \kappa^{-1}$, the canonically normalized scalar field $\phi$ is defined by
\begin{equation}
\phi \equiv \sqrt{\frac{3}{2}}\,M_{\textrm Pl}\,\ln f_R(R).
\label{eq:phi_def}
\end{equation}
\begin{figure}[t]
  \centering
  \subfigure[~Einstein-frame scalaron potential $V(\phi)$.]{\label{fig:9}\includegraphics[width=0.25\textwidth]{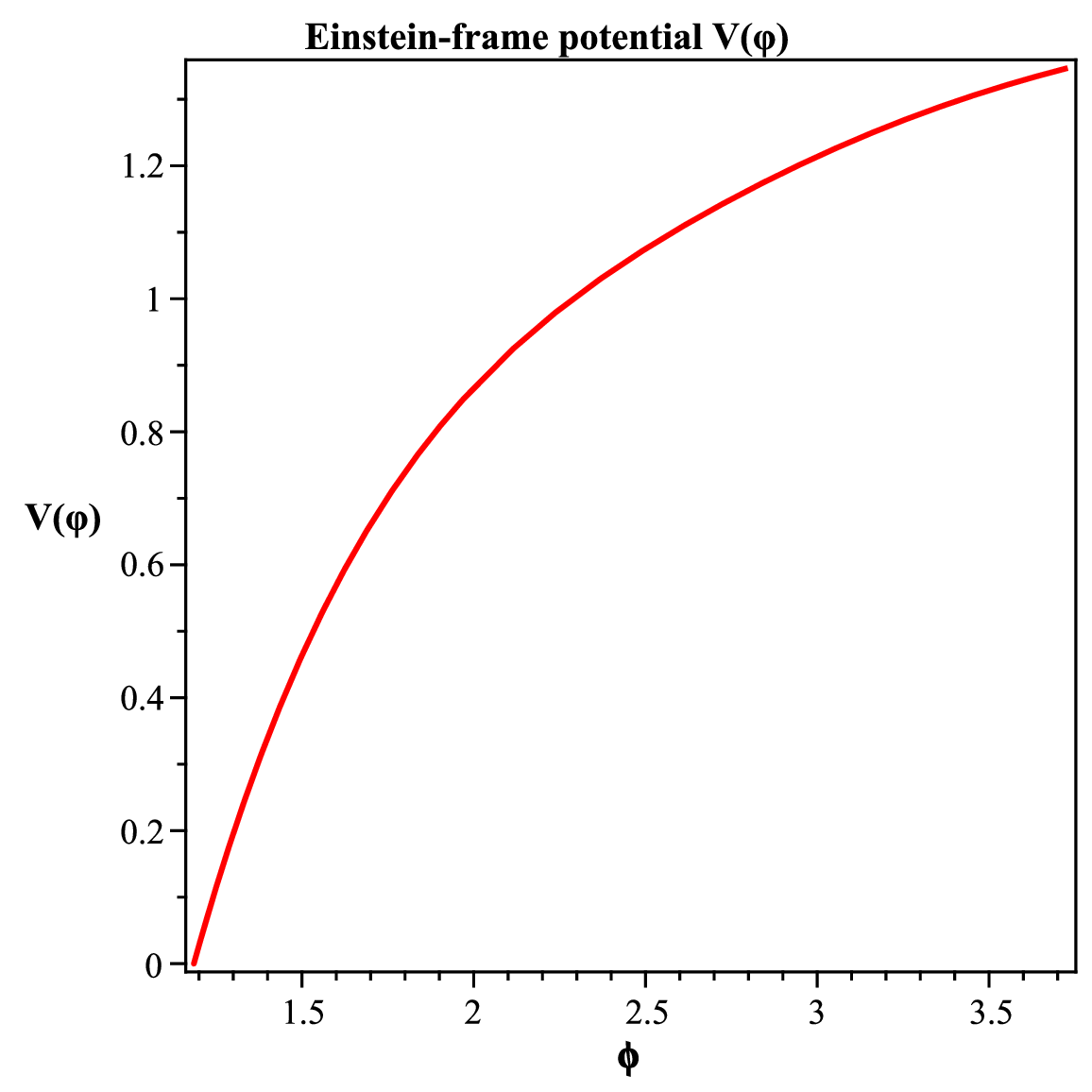}}
  \subfigure[~Time evolution of the scalaron field $\phi(t)$.]{\label{fig:10}\includegraphics[width=0.25\textwidth]{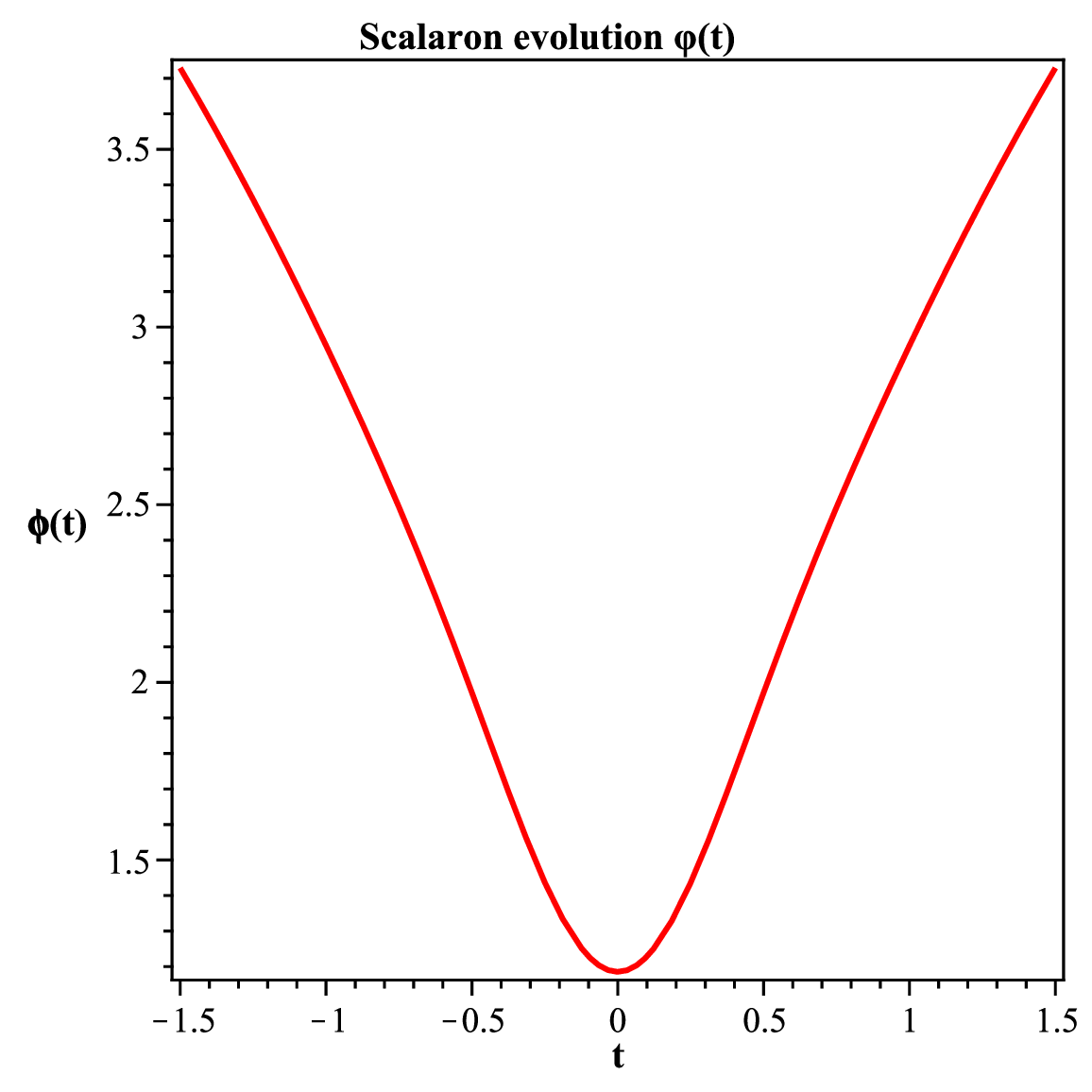}}
  \caption{Einstein-frame description of the vacuum bounce for the extended model
$f(R)=R-2\Lambda+\alpha R^{2}\!\left(1-e^{-R/R_b}\right)$.
Panel \subref{fig:9} shows the scalaron potential
$V(\phi)=\frac{M_{\textrm Pl}^2}{2}\,(R f_R-f)/f_R^{2}$ as a function of the canonically
normalized scalar field $\phi=\sqrt{3/2}\,M_{\textrm Pl}\ln f_R$.
Panel \subref{fig:10} displays the time evolution of $\phi(t)$ along the Jordan-frame bouncing
background.
The scalar field remains finite and evolves smoothly across the bounce,
reaching a turning point near $t=0$, which signals the occurrence of the bounce
in the Einstein frame.
}
\label{fig:einstein_frame}
\end{figure}
Figure~\ref{fig:einstein_frame} provides a transparent physical interpretation of
the vacuum bounce in terms of the Einstein-frame scalar degree of freedom.
As shown in Fig.~\ref{fig:einstein_frame}\subref{fig:9}, the scalaron potential exhibits a smooth, monotonic
structure without singularities, ensuring well-defined scalar-field dynamics.
The evolution of $\phi(t)$ in Fig.~\ref{fig:einstein_frame}\subref{fig:10} confirms that the scalar field reaches a
turning point at the bounce, where $\dot{\phi}=0$ while $\phi$ itself remains finite.

With this definition, the gravitational action becomes the Einstein-Hilbert action plus a canonical scalar field,
\begin{equation}
S=\int d^4x\,\sqrt{-\tilde g}\left[\frac{M_{\textrm Pl}^2}{2}\tilde R
-\frac{1}{2}\tilde g^{\mu\nu}\partial_\mu\phi\,\partial_\nu\phi
-V(\phi)\right],
\label{eq:Einstein_action}
\end{equation}
where $\tilde R$ is the Ricci scalar of the Einstein-frame metric $\tilde g_{\mu\nu}$.

The scalaron potential is determined by the Jordan-frame function $f(R)$ through
\begin{equation}
V(\phi(R))=\frac{M_{\textrm Pl}^2}{2}\,\frac{R f_R(R)-f(R)}{f_R(R)^2}.
\label{eq:V_of_R}
\end{equation}
Substituting Eq.~\eqref{eq:fR_minus_f_ext} into Eq.~\eqref{eq:V_of_R}, the potential becomes
\begin{equation}
V(\phi(R))
=\frac{M_{\textrm Pl}^2}{2}\,
\frac{
2\Lambda+\alpha\left[
R^2\left(1-e^{-R/R_b}\right)
+\frac{R^3}{R_b}e^{-R/R_b}
\right]
}{
\left[
1+\alpha\left(
2R\left(1-e^{-R/R_b}\right)+\frac{R^{2}}{R_b}e^{-R/R_b}
\right)
\right]^2
}.
\label{eq:V_explicit}
\end{equation}
In practice, for numerical work one evaluates $V$ as a function of time $t$ through $R(t)$, and constructs $\phi(t)$ via Eq.~\eqref{eq:phi_def}. The pair $(\phi(t),V(t))$ can then be interpolated to obtain $V(\phi)$ if desired.


Under the conformal transformation~\eqref{eq:conformal_transform}, the cosmic times in the two frames are related by
\begin{equation}
d\tilde t = \sqrt{f_R(R(t))}\, dt, \quad \mbox{and the scale factors are related by} \quad
\tilde a(\tilde t)=\sqrt{f_R(R(t))}\,a(t).
\label{eq:scale_factor_relation}
\end{equation}
Consequently, the Einstein-frame Hubble parameter is
\begin{equation}
\tilde H \equiv \frac{1}{\tilde a}\frac{d\tilde a}{d\tilde t}, \quad \mbox{using Eq.~\eqref{eq:scale_factor_relation} one finds the practical Jordan-frame expression}\quad
\tilde H(t)=\frac{1}{\sqrt{f_R}}\left(H+\frac{\dot f_R}{2f_R}\right).
\label{eq:Htilde_in_t}
\end{equation}
Equation~\eqref{eq:Htilde_in_t} is useful because $H(t)$ and $f_R(t)$ can be computed directly from the Jordan-frame bounce background.


From Eq.~\eqref{eq:phi_def}, the time derivative of $\phi$ with respect to Jordan time is \cite{ElHanafy:2014efn}
\begin{equation}
\dot\phi(t)=\sqrt{\frac{3}{2}}\,M_{\textrm Pl}\,\frac{\dot f_R}{f_R}.
\label{eq:phidot}
\end{equation}
The derivative with respect to Einstein-frame time $\tilde t$ is therefore
\begin{equation}
\phi'(\tilde t)\equiv \frac{d\phi}{d\tilde t}
=\frac{\dot\phi}{d\tilde t/dt}
=\sqrt{\frac{3}{2}}\,M_{\textrm Pl}\,\frac{\dot f_R}{f_R^{3/2}}.
\label{eq:phi_prime}
\end{equation}
In numerical implementations, $\dot f_R$ can be evaluated using
\begin{equation}
\dot f_R = f_{RR}\,\dot R, \quad \mbox{and from Eq.~\eqref{eq:Ricci_bounce} one has} \quad
R(t)=R_0+48\lambda^2 t^2,
\qquad
\dot R(t)=96\lambda^2 t.
\label{eq:R_and_Rdot}
\end{equation}

Given a viable parameter set $(\alpha,R_b)$ (and $\Lambda$ fixed by the bounce condition), the Einstein-frame background required for perturbation evolution is constructed by the following steps:
\begin{enumerate}
\item Compute $R(t)$ and $\dot R(t)$ from Eq.~\eqref{eq:R_and_Rdot}.
\item Evaluate $f_R(R(t))$ and $f_{RR}(R(t))$ from Eqs.~\eqref{eq:fR_boxed}--\eqref{eq:fRR_boxed}.
\item Construct the Einstein-frame time grid via numerical integration of Eq.~\eqref{eq:scale_factor_relation}.
\item Obtain $\tilde a(t)$ from Eq.~\eqref{eq:scale_factor_relation} and $\tilde H(t)$ from Eq.~\eqref{eq:Htilde_in_t}.
\item Construct $\phi(t)$ using Eq.~\eqref{eq:phi_def} and $\phi'(\tilde t)$ using Eq.~\eqref{eq:phi_prime}.
\item Compute the potential $V(t)$ via Eq.~\eqref{eq:V_explicit}.
\end{enumerate}
These background quantities $(\tilde a,\tilde H,\phi,\phi',V)$ provide the complete input needed to evolve tensor and scalar perturbations in the Einstein frame, which is the subject of the next section.

\section{Cosmological Perturbations Across the Bounce}
\label{sec:perturbations}

In this section we present the equations governing tensor and scalar perturbations in the Einstein frame and outline the numerical strategy used to evolve them smoothly across the bounce. Working in the Einstein frame allows us to treat the system as GR coupled to a canonical scalar field, avoiding spurious singularities that may arise in the Jordan-frame formulation.

\subsection{Tensor perturbations}

Tensor perturbations describe primordial gravitational waves and are gauge-invariant at linear order. In the Einstein frame, the perturbed metric including tensor modes can be written as
\begin{equation}
d\tilde s^2
=-d\tilde t^2
+\tilde a^2(\tilde t)\left(\delta_{ij}+\tilde h_{ij}\right)dx^i dx^j,
\end{equation}
where $\tilde h_{ij}$ is transverse and traceless:
\begin{equation}
\partial^i \tilde h_{ij}=0,
\qquad
\tilde h^i{}_i=0.
\end{equation}

Expanding $\tilde h_{ij}$ in Fourier modes and polarizations,
\begin{equation}
\tilde h_{ij}(\tilde t,\vec x)
=\sum_{\lambda=+,\times}\int\frac{d^3k}{(2\pi)^3}
\,\epsilon^{(\lambda)}_{ij}(\vec k)\,
\tilde h_k^{(\lambda)}(\tilde t)\,
e^{i\vec k\cdot\vec x}, \quad \mbox{where each mode obeys the equation of motion} \quad
\tilde h_k''+3\tilde H\,\tilde h_k'
+\frac{k^2}{\tilde a^2}\tilde h_k=0,
\label{eq:tensor_eq}
\end{equation}
where a prime denotes a derivative with respect to Einstein-frame time $\tilde t$.

\paragraph{Regularity across the bounce.}
Since $\tilde a(\tilde t)$ and $\tilde H(\tilde t)$ are finite and continuous for viable models with $f_R>0$, Eq.~\eqref{eq:tensor_eq} has no singular coefficients at the bounce. The absence of divergences in $\tilde h_k$ therefore provides a direct test of the stability of tensor perturbations.

\begin{figure}[t]
  \centering
  \subfigure[~Tensor friction term $\Gamma(t)=3\tilde H(t)$ in the Einstein frame.]{\label{fig:11}\includegraphics[width=0.25\textwidth]{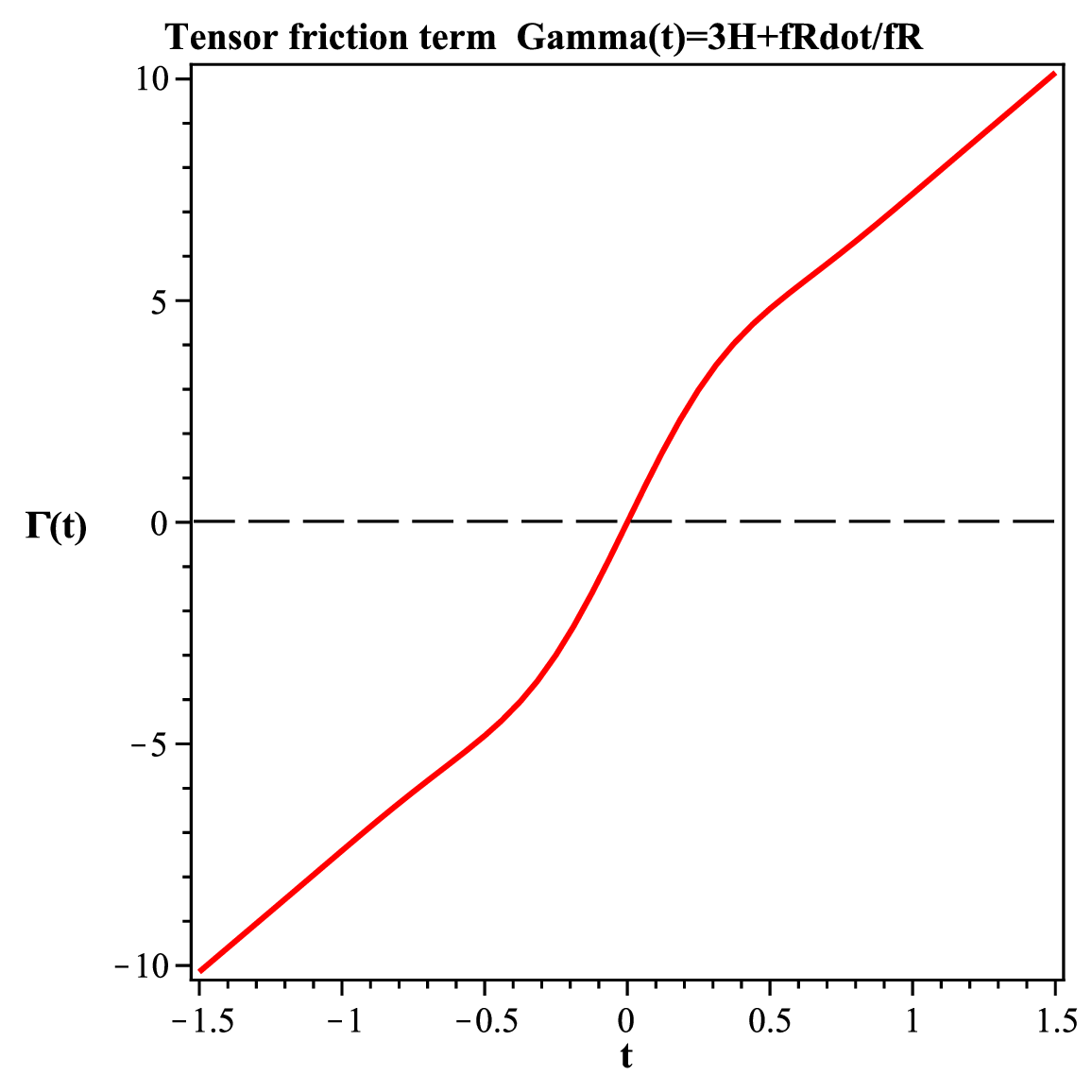}}
  \subfigure[~Evolution of tensor perturbation modes $\tilde h_k(\tilde t)$ across the bounce.]{\label{fig:12}\includegraphics[width=0.25\textwidth]{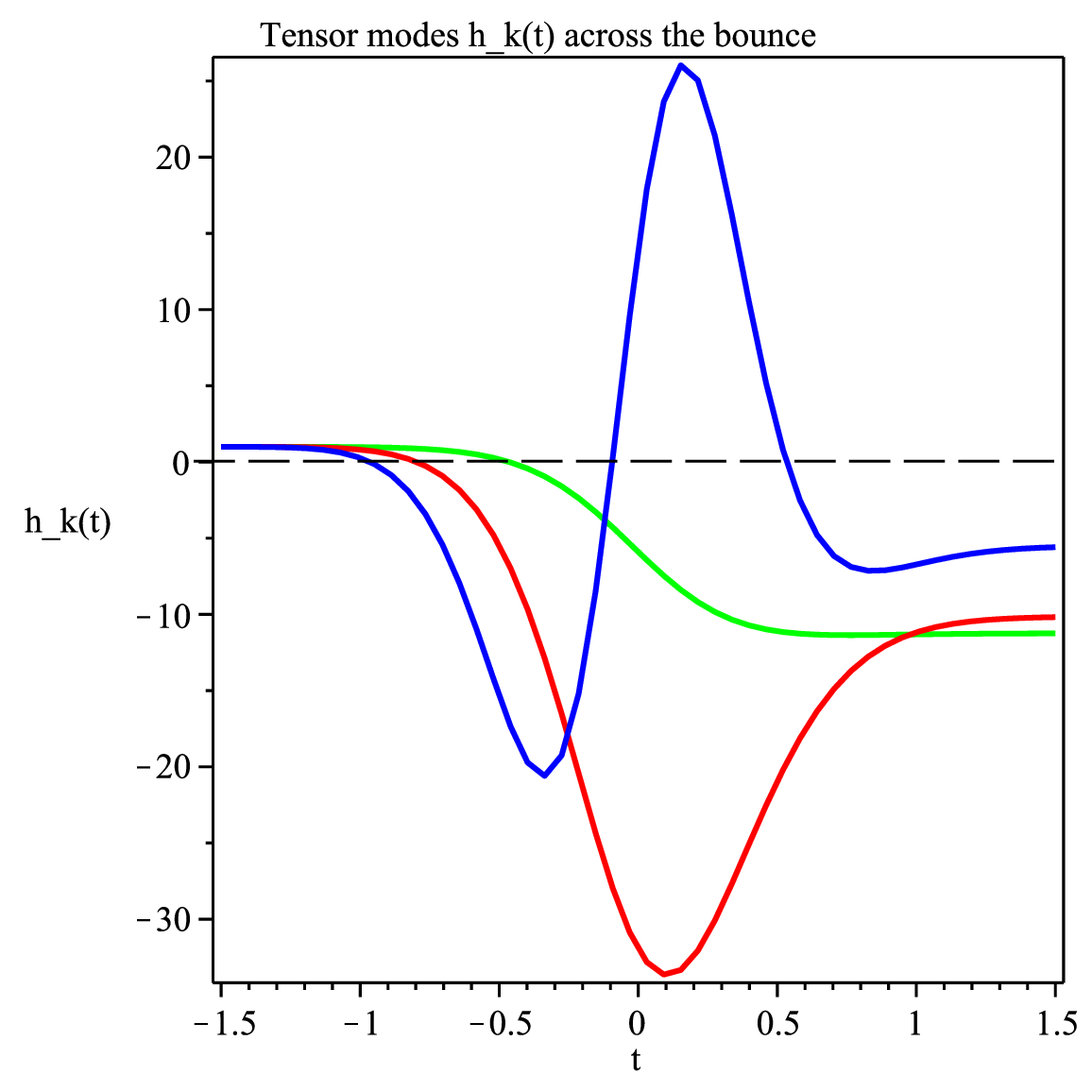}}
 \caption{Tensor perturbations across the vacuum bounce in the Einstein frame.
Panel \subref{fig:11} shows the effective friction term $\Gamma(t)=3\tilde H(t)$ governing the
evolution of tensor modes, computed from the background quantities constructed
in Sec.~\ref{sec:perturbations}. The friction term remains finite and continuous throughout the bounce.
Panel \subref{fig:12} displays the numerical evolution of tensor perturbation modes
$\tilde h_k(\tilde t)$ for several representative comoving wavenumbers $k$.
All modes remain finite and well behaved across the bounce, with no evidence of
divergent or exponentially growing behavior.
}
\label{fig:tensor_perturbations}
\end{figure}
Figure~\ref{fig:tensor_perturbations} demonstrates the stability of tensor
perturbations across the nonsingular vacuum bounce.
As shown in Fig.~\ref{fig:tensor_perturbations}~\subref{fig:11}, the effective friction term $\Gamma(t)=3\tilde H(t)$ remains
finite and smoothly varying through the bounce, indicating the absence of
singular behavior in the tensor evolution equation.
The numerical solutions displayed in Fig.~\ref{fig:tensor_perturbations}~\subref{fig:12} confirm that tensor perturbation
modes $\tilde h_k(\tilde t)$ propagate smoothly from the contracting to the
expanding phase without divergences or pathological growth.

Scalar perturbations are most conveniently described using the Mukhanov--Sasaki variable $v$, which is gauge-invariant and captures the dynamics of curvature perturbations. In the Einstein frame, the action for scalar perturbations reduces to
\begin{equation}
S^{(2)}=\frac{1}{2}\int d\tilde t\,d^3x
\left[
(v')^2-(\nabla v)^2+\frac{z''}{z}v^2
\right], \quad \mbox{where} \quad
z(\tilde t)\equiv \frac{\tilde a\,\phi'}{\tilde H}.
\label{eq:z_def}
\end{equation}

Varying the action yields the Mukhanov--Sasaki equation for each Fourier mode:
\begin{equation}
v_k''+\left(
\frac{k^2}{\tilde a^2}-\frac{z''}{z}
\right)v_k=0, \quad \mbox{where the comoving curvature perturbation is given by} \quad
\mathcal{R}_k=\frac{v_k}{z}.
\label{eq:curvature_perturbation}
\end{equation}

\begin{figure}[t]
  \centering
  \subfigure[~Time evolution of the Ricci scalar $R(t)$ near the bounce.]{\label{fig:11}\includegraphics[width=0.25\textwidth]{JBFMMM_ISI_R}}
  \subfigure[~Evolution of Mukhanov--Sasaki scalar perturbation modes $v_k(t)$ across the bounce.]{\label{fig:12}\includegraphics[width=0.25\textwidth]{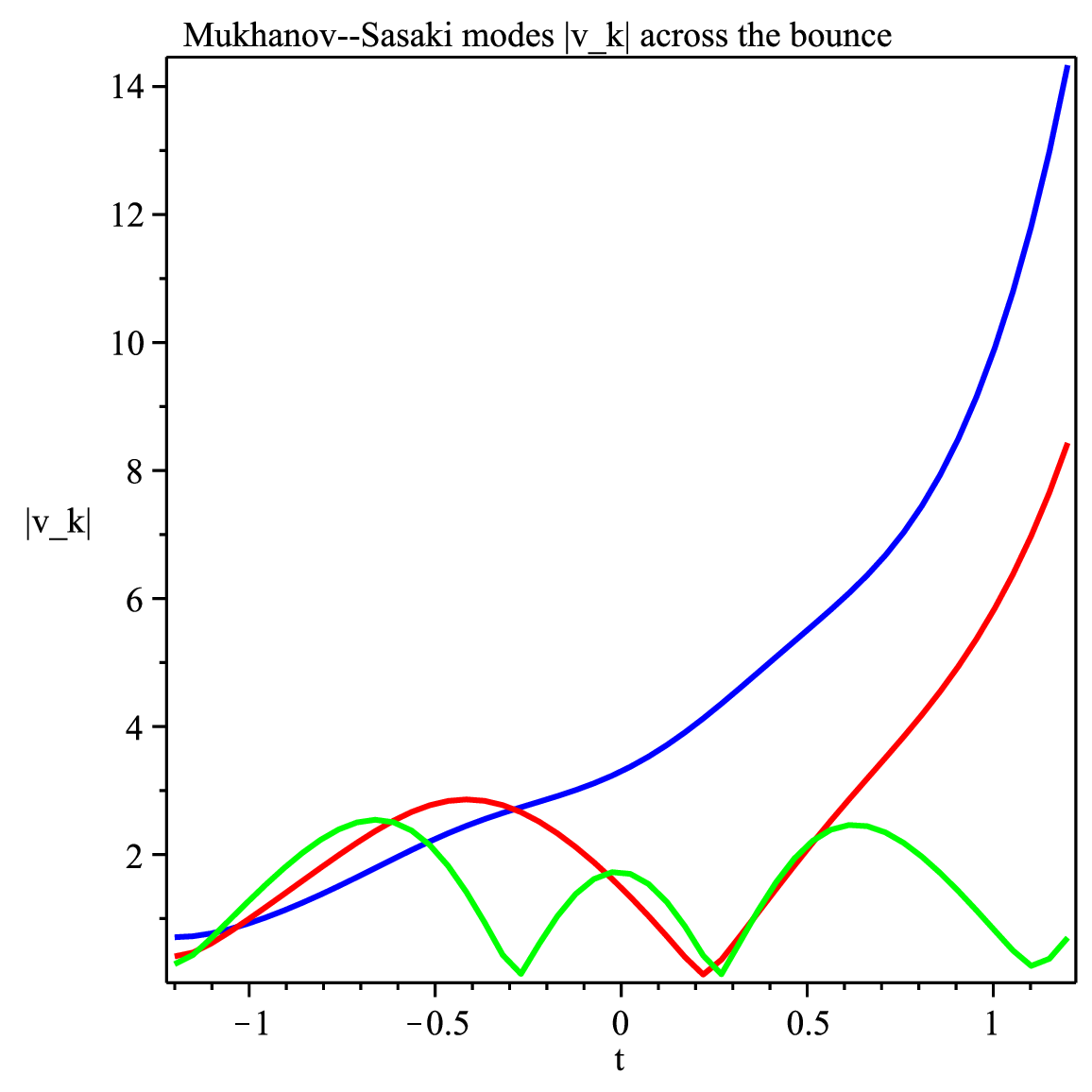}}
    \caption{Behavior of background curvature and scalar perturbations across the nonsingular vacuum bounce.
    Panel \subref{fig:11} shows the time evolution of the Ricci scalar $R(t)$, which remains finite and reaches a smooth
    minimum at the bounce point $t=0$, confirming the absence of curvature singularities.
    Panel \subref{fig:11} displays the numerical evolution of the Mukhanov--Sasaki variable $v_k(t)$ for representative
    comoving wavenumbers. All modes remain finite and well behaved throughout the transition from contraction
    to expansion, indicating the stability of scalar perturbations across the bounce.}
    \label{fig:scalar-bounce}
\end{figure}

Figure~\ref{fig:scalar-bounce} provides a combined test of the background dynamics and the scalar perturbation
sector of the model. The left panel demonstrates that the Ricci scalar evolves smoothly and symmetrically
around the bounce, attaining a finite minimum value at $t=0$. This confirms that the bouncing solution is
genuinely nonsingular at the curvature level. The right panel shows that the Mukhanov--Sasaki variable $v_k$, which governs gauge-invariant scalar
perturbations in the Einstein frame, evolves regularly across the bounce for all displayed modes. No
divergences or pathological growth are observed near the bounce point. This behavior indicates that the
effective potential $z''/z$ remains regular and that scalar perturbations can be consistently propagated
from the contracting phase to the expanding phase.

\newpage
In bouncing cosmologies, special care is required near the bounce where $\tilde H$ may approach zero. In the present construction:
\begin{itemize}
\item $\tilde a(\tilde t)$ remains finite and nonzero,
\item $\phi'(\tilde t)$ is finite for viable parameter choices,
\item $z(\tilde t)$ remains finite provided $\tilde H$ does not vanish faster than $\phi'$.
\end{itemize}
Numerically, we monitor $z$ and $z''/z$ explicitly to ensure that no spurious divergences arise. The regularity of $\mathcal{R}_k$ across the bounce is taken as the primary criterion for scalar perturbation stability.


Initial conditions for perturbations are imposed in the far pre-bounce phase, $\tilde t=\tilde t_i\ll 0$, where all relevant modes are deep inside the horizon,
\begin{equation}
\frac{k}{\tilde a(\tilde t_i)\tilde H(\tilde t_i)}\gg 1.
\end{equation}
In this regime, the modes behave as in Minkowski spacetime, and we adopt the Bunch--Davies vacuum:
\begin{align}
v_k(\tilde t_i) = \frac{1}{\sqrt{2k}}, \qquad
v_k'(\tilde t_i) = -i\sqrt{\frac{k}{2}}, \qquad
\tilde h_k(\tilde t_i) = \frac{1}{\sqrt{2k}}, \qquad
\tilde h_k'(\tilde t_i) = -i\sqrt{\frac{k}{2}}.
\end{align}
These conditions ensure a well-defined quantum vacuum for both scalar and tensor perturbations.


Equations~\eqref{eq:tensor_eq} and~\eqref{eq:curvature_perturbation} are evolved numerically across the bounce for a range of comoving wavenumbers $k$. A viable bouncing solution is required to satisfy:
\begin{enumerate}
\item Finite and continuous tensor modes $\tilde h_k(\tilde t)$ across the bounce,
\item Finite Mukhanov--Sasaki variable $v_k(\tilde t)$,
\item Regular comoving curvature perturbation $\mathcal{R}_k(\tilde t)$ with no divergences or unphysical growth.
\end{enumerate}

Successful fulfillment of these conditions indicates that the bounce is not only geometrically consistent but also dynamically stable at the level of linear perturbations.

In addition to remaining finite, the scalar perturbation modes exhibit no significant amplification
or mode mixing as they propagate through the bounce. The Mukhanov--Sasaki variable $v_k$ evolves
smoothly for all wavenumbers considered, and its amplitude remains within the linear regime
throughout the transition from contraction to expansion.
This confirms that the stability observed in Fig.~\ref{fig:scalar-bounce} is not merely qualitative, but also ensures
the self-consistency of the linear perturbation analysis across the bounce.

Quantitatively, the amplitudes of both scalar and tensor modes remain of the same order of magnitude
across the bounce, with no indication of large amplification that would signal a breakdown of the
linear perturbative regime.

In the following section, we will present representative numerical results illustrating stable tensor and scalar mode evolution for selected points in the viable parameter space.

\section{Numerical Results}
\label{sec:results}

In this section we summarize the main numerical findings of the parameter-space scan and the perturbation analysis. The emphasis is on identifying robust qualitative features rather than exhaustive phenomenological constraints.

\subsection{Viable parameter regions}

The parameter scan in the $(\bar\alpha,\bar R_b)$ plane reveals extended regions where the vacuum bounce is both geometrically realized and dynamically viable. Specifically, we find that:
\begin{itemize}
\item For moderate positive values of $\bar R_b=\mathcal{O}(1)$, there exists a continuous range of $\bar\alpha>0$ for which \cite{Nashed:2021pah}
\[
f_R(\bar R(t))>0,
\qquad
f_{RR}(\bar R(t))>0, \quad \mbox{throughout the interval \quad $t\in[-T,T]$}. \]
\item Extremely small values of $\bar R_b$ tend to produce rapid variations in $f_R$ and $f_{RR}$, leading to violations of the tachyon-free condition near the bounce.
\item Large values of $\bar R_b$ suppress the exponential term, effectively approaching the pure $R+\alpha R^2$ limit, in which the viable region shrinks unless $\bar\alpha$ is sufficiently small.
\end{itemize}

For each viable point in parameter space, the constant $\Lambda$ determined by Eq.~\eqref{eq:Lambda_dimensionless} ensures the exact satisfaction of the vacuum bounce condition.


Tensor modes were evolved numerically using Eq.~\eqref{eq:tensor_eq} for a representative set of viable parameter choices. In all cases examined:
\begin{itemize}
\item The Einstein-frame scale factor $\tilde a(\tilde t)$ remains finite and nonvanishing across the bounce.
\item Tensor amplitudes $\tilde h_k(\tilde t)$ remain finite and continuous for all modes considered.
\item No exponential growth or instability is observed near the bounce point.
\end{itemize}
These results indicate that the bounce is stable with respect to primordial gravitational-wave perturbations.


Scalar perturbations were evolved using the Mukhanov--Sasaki equation~\eqref{eq:curvature_perturbation}. For viable parameter choices:
\begin{itemize}
\item The function $z(\tilde t)$ defined in Eq.~\eqref{eq:z_def} remains finite across the bounce.
\item The effective potential $z''/z$ exhibits a smooth, localized structure centered around the bounce, without singular behavior.
\item The Mukhanov--Sasaki variable $v_k(\tilde t)$ and the comoving curvature perturbation $\mathcal{R}_k(\tilde t)$ remain finite and well behaved across the bounce.
\end{itemize}
This demonstrates that scalar perturbations can be transferred smoothly from the contracting to the expanding phase in the present framework.

{  We emphasize that the perturbation analysis presented here is intended to establish the internal consistency and linear stability of the bouncing background rather than to provide detailed phenomenological predictions for the primordial power spectrum. Our primary goal is to demonstrate that tensor and scalar perturbations evolve smoothly and remain finite across the bounce, and that no ghost or tachyonic instabilities arise within the viable parameter region. In addition to remaining finite, the scalar and tensor perturbation modes exhibit no significant amplification or pathological mode mixing as they propagate through the bounce for the representative range of comoving wavenumbers considered in our numerical runs.  A full derivation of observable quantities such as spectral indices, tensor-to-scalar ratios, or detailed matching to a post-bounce radiation-dominated phase would require additional assumptions about the pre-bounce vacuum state, matter content, and reheating dynamics. Such a phenomenological analysis lies beyond the scope of the present work and is deferred to future investigation.
}

\section{Conclusions and discussion}
\label{sec:conclusions}

In this work we have constructed and analyzed a minimal and controlled realization of a nonsingular cosmological bounce in metric $f(R)$ gravity. Our main results can be summarized as follows:
\begin{enumerate}
\item We adopted a smooth Gaussian-type bouncing scale factor and showed that the corresponding curvature is finite and positive at the bounce.
\item For the exponential switch-on model
\[
f(R)=R+\alpha R^2\left(1-e^{-R/R_b}\right),
\]
we derived a no-go result demonstrating that a positive-curvature vacuum bounce cannot be supported without further modification within metric $f(R)$ gravity in vacuum and for $R_0$ > 0.
\item Introducing a minimal constant term $-2\Lambda$ removes this obstruction and allows the bounce condition to be satisfied exactly, with $\Lambda$ fixed algebraically by the bounce curvature.
\item A systematic parameter scan identified regions in which the theory remains ghost-free and tachyon-free throughout the bouncing phase.
\item By working in the Einstein frame, we showed that both tensor and scalar perturbations evolve smoothly across the bounce, with no divergences or instabilities.
\end{enumerate}

The framework developed here provides a clean and technically consistent example of a geometrically induced bounce in $f(R)$ gravity that goes beyond background-level considerations. Future work may extend this analysis by including realistic matter components, studying the generation of primordial spectra, or connecting the post-bounce evolution to standard radiation- and matter-dominated cosmology.

Overall, our results demonstrate that viable and perturbatively stable bouncing cosmologies can be realized within a minimal extension of $f(R)$ gravity, reinforcing its role as a promising framework for nonsingular early-universe scenarios.

Working in the Einstein frame provides a clear and intuitive interpretation of the vacuum bounce.
The conformal transformation maps the $f(R)$ theory to GR coupled to a canonical scalar
field (the scalaron), whose dynamics remain regular across the bounce for all viable parameter choices.
The scalaron potential is smooth and free of singularities, and the scalar field reaches a turning
point at the bounce while remaining finite, indicating that the bounce corresponds to a controlled
reversal in field-space rather than a pathological breakdown of the theory.

The numerical evolution of tensor and scalar perturbations further confirms the robustness of the
bounce. Tensor modes propagate smoothly across the bounce, with finite amplitudes and no evidence
of divergent or exponentially growing behavior. Similarly, the Mukhanov--Sasaki variable and the
comoving curvature perturbation remain regular for all modes considered.
The absence of singularities in the effective friction terms and perturbation potentials demonstrates
that the bounce is dynamically stable at the level of linear perturbations.

Compared to matter-bounce models or scenarios requiring exotic fluids, the vacuum bounce realized
here is driven entirely by geometric effects.
The minimal extension proposed in this work avoids introducing additional matter components or
noncanonical kinetic terms, thereby preserving the conservative nature of metric $f(R)$ gravity.
At the same time, the matter-bounce variant discussed in Appendix~A illustrates that the same
exponential switch-on model can support a phenomenologically richer scenario when matter is included,
providing a direct link to scale-invariant perturbation generation.

Table~\ref{tab:comparison} summarizes the essential differences between the previously studied exponential switch-on $f(R)$ model and the present extended framework developed in this work. While the pure exponential deformation of the Starobinsky model fails to satisfy the exact vacuum bounce condition at positive curvature, leading to a no-go result, the present study demonstrates that this obstruction can be removed through a minimal extension by a constant term. Importantly, this modification does not introduce additional degrees of freedom, nor does it alter the high-curvature behavior or the ghost- and tachyon-free structure of the theory. Instead, the constant $\Lambda$ is fixed uniquely by the bounce condition itself, allowing for an exact realization of a nonsingular vacuum bounce. Furthermore, working in the Einstein frame shows that both scalar and tensor perturbations remain finite and well behaved across the bounce, establishing the present construction as a minimal and perturbatively stable realization of a geometric vacuum bounce in $f(R)$ gravity.
\clearpage

\begin{table}[t]
\centering
\caption{Comparison between the previous exponential switch-on $f(R)$ model and the present extended model for realizing a vacuum cosmological bounce.}
\label{tab:comparison}

\scriptsize
\renewcommand{\arraystretch}{1.7}

\begin{tabular}{l @{\hspace{0.3cm}} l @{\hspace{0.3cm}} l}
\hline
\textbf{Feature} & \textbf{Previous study} & \textbf{Present study} \\
\hline

Gravitational action 
& $f(R)=R+\alpha R^{2}(1-e^{-R/R_{b}})$ 
& $f(R)=R-2\Lambda+\alpha R^{2}(1-e^{-R/R_{b}})$ \\

Vacuum bounce realization 
& Not possible for $R_{0}>0$ (no-go result) 
& Exactly realized for $R_{0}>0$ \\

Bounce condition 
& $f(R_{0}) \neq R_{0}f_{R}(R_{0})$ unless $\alpha=0$ 
& $f(R_{0}) = R_{0}f_{R}(R_{0})$ with $\Lambda$ fixed algebraically \\

Role of additional parameters 
& No compensating mechanism available 
& Constant $\Lambda$ compensates the geometric obstruction \\

Number of degrees of freedom 
& One extra scalar degree of freedom (scalaron) 
& Same (no new degrees of freedom introduced) \\

Ghost-free condition ($f_{R}>0$) 
& Not sufficient to ensure a bounce 
& Satisfied in viable parameter regions \\

Tachyon-free condition ($f_{RR}>0$) 
& Does not rescue the vacuum bounce 
& Ensures perturbative stability around the bounce \\

High-curvature behavior 
& Approaches Starobinsky $R+\alpha R^{2}$ 
& Unchanged compared to the previous model \\

Einstein-frame dynamics 
& Bounce obstructed at background level 
& Smooth scalaron evolution with finite potential \\

Scalar and tensor perturbations 
& Not consistently supported across the bounce 
& Remain finite and well behaved across the bounce \\

Physical outcome 
& Geometric vacuum bounce forbidden 
& Minimal and stable geometric vacuum bounce achieved \\

\hline
\end{tabular}
\end{table}
\subsection{Outlook}

The framework developed in this study provides a clean benchmark for constructing nonsingular
cosmological models in higher-curvature gravity.
Future work may extend the present analysis by incorporating realistic matter sectors,
studying the generation and observational properties of primordial spectra,
or connecting the post-bounce evolution to standard radiation- and matter-dominated cosmology.
Overall, the results demonstrate that viable and perturbatively stable vacuum bounces can be
realized within a minimal extension of $f(R)$ gravity, reinforcing its role as a promising
framework for nonsingular early-universe scenarios.

\appendix
\section{Matter Bounce Variant}
\label{app:matter_bounce}

In this appendix we briefly extend the analysis of the main text by allowing for a matter component at the bounce. This provides a complementary realization of the bounce mechanism and allows contact with the well-known matter-bounce scenario, while keeping the same $f(R)$ framework.

\subsection{Matter bounce background}

Instead of the Gaussian bounce used in the main text, we consider a scale factor that asymptotically mimics a matter-dominated universe far from the bounce,
\begin{equation}
a(t)=a_b\left(1+\frac{t^2}{t_0^2}\right)^{1/3},
\label{eq:matter_bounce_scale_factor}
\end{equation}
where $a_b>0$ and $t_0>0$ are constants. The corresponding Hubble parameter is
\begin{equation}
H(t)=\frac{2t}{3(t^2+t_0^2)}, \quad \mbox{which satisfies} \quad
H(0)=0, \qquad \dot H(0)=\frac{2}{3t_0^2}>0,
\end{equation}
confirming the existence of a nonsingular bounce at $t=0$.

The Ricci scalar for this background is
\begin{equation}
R(t)=6\left(\dot H+2H^2\right)
=\frac{4(3t_0^2- t^2)}{3(t^2+t_0^2)^2},
\label{eq:matter_bounce_R}
\end{equation}
which remains finite throughout the evolution and approaches $R\rightarrow 0$ for $|t|\gg t_0$, consistent with a matter-dominated phase.

\subsection{Modified Friedmann equation with matter}

Allowing for a matter component with energy density $\rho$, the modified Friedmann equation in $f(R)$ gravity reads
\begin{equation}
3H^2 f_R
=\frac{1}{2}\left(f_R R-f\right)
-3H\dot f_R
+\kappa^2\rho.
\end{equation}
At the bounce, where $H=0$, this reduces to
\begin{equation}
\frac{1}{2}\left[f_R(R_0)R_0-f(R_0)\right]
+\kappa^2\rho_b=0,
\label{eq:matter_bounce_condition}
\end{equation}
with $R_0=R(0)=4/(t_0^2)$ and $\rho_b$ the matter energy density at the bounce.

Equation~\eqref{eq:matter_bounce_condition} shows that, unlike the vacuum case, the geometric contribution need not vanish by itself. Instead, it can be balanced by a positive matter energy density.

\subsection*{Acknowledgments}
This work was supported and funded by the Deanship of Scientific Research at Imam Mohammad Ibn Saud Islamic University (IMSIU) (grant number IMSIU-DDRSP2602).
%

\end{document}